\PassOptionsToPackage{unicode}{hyperref}
\PassOptionsToPackage{hyphens}{url}
\PassOptionsToPackage{dvipsnames,svgnames,x11names}{xcolor}
\documentclass[
  11pt,
]{article}
\usepackage{xcolor}
\usepackage[margin=1in]{geometry}
\usepackage{amsmath,amssymb}
\usepackage{iftex}
\ifPDFTeX
  \usepackage[T1]{fontenc}
  \usepackage[utf8]{inputenc}
  \usepackage{textcomp} 
\else 
  \usepackage{unicode-math} 
  \defaultfontfeatures{Scale=MatchLowercase}
  \defaultfontfeatures[\rmfamily]{Ligatures=TeX,Scale=1}
\fi
\usepackage{lmodern}
\IfFileExists{upquote.sty}{\usepackage{upquote}}{}
\IfFileExists{microtype.sty}{
  \usepackage[]{microtype}
  \UseMicrotypeSet[protrusion]{basicmath} 
}{}
\makeatletter
\@ifundefined{KOMAClassName}{
  \IfFileExists{parskip.sty}{%
    \usepackage{parskip}
  }{
    \setlength{\parindent}{0pt}
    \setlength{\parskip}{6pt plus 2pt minus 1pt}}
}{
  \KOMAoptions{parskip=half}}
\makeatother
\usepackage{longtable,booktabs,array}
\usepackage{calc} 
\usepackage{etoolbox}
\makeatletter
\patchcmd\longtable{\par}{\if@noskipsec\mbox{}\fi\par}{}{}
\makeatother
\IfFileExists{footnotehyper.sty}{\usepackage{footnotehyper}}{\usepackage{footnote}}
\makesavenoteenv{longtable}
\usepackage{graphicx}
\makeatletter
\newsavebox\pandoc@box
\newcommand*\pandocbounded[1]{
  \sbox\pandoc@box{#1}%
  \Gscale@div\@tempa{\textheight}{\dimexpr\ht\pandoc@box+\dp\pandoc@box\relax}%
  \Gscale@div\@tempb{\linewidth}{\wd\pandoc@box}%
  \ifdim\@tempb\p@<\@tempa\p@\let\@tempa\@tempb\fi
  \ifdim\@tempa\p@<\p@\scalebox{\@tempa}{\usebox\pandoc@box}%
  \else\usebox{\pandoc@box}%
  \fi%
}
\def\fps@figure{htbp}
\makeatother
\providecommand{\tightlist}{%
  \setlength{\itemsep}{0pt}\setlength{\parskip}{0pt}}
\usepackage[]{natbib}
\usepackage{mathtools}
\usepackage{enumitem}
\usepackage{caption}
\setlist{nosep}
\AtBeginEnvironment{longtable}{\small}
\usepackage{bookmark}
\IfFileExists{xurl.sty}{\usepackage{xurl}}{} 
\makeatletter
\@ifundefined{xmpquote}{}{}
\makeatother
\hypersetup{
  pdftitle={Who Can Make the Action Happen? An Authority-Decomposition Framework for High-Risk Automated Systems},
  pdfauthor={Mengting Wu; Lin Wang; Yong Zhang},
  colorlinks=true,
  linkcolor={blue},
  filecolor={Maroon},
  citecolor={blue},
  urlcolor={blue},
  pdfcreator={LaTeX via pandoc}}

\title{Who Can Make the Action Happen? An Authority-Decomposition
Framework for High-Risk Automated Systems}
\author{Mengting Wu\textsuperscript{*}, Lin Wang, and Yong Zhang\\
\small Chengdu Havenlon Security Technology Co., Ltd., Chengdu, China\\
\small \textsuperscript{*}Corresponding author: \texttt{chloe@havenlon.com}}
\date{}

\begin{document}
\maketitle

\begin{center}
\small\textbf{Status:} Non-peer-reviewed preprint for scholarly feedback.
\end{center}

\section*{Abstract}\label{abstract}
\addcontentsline{toc}{section}{Abstract}

High-risk automated systems distribute control across services,
credentials, protected components, and lifecycle mechanisms. Labels such
as \emph{authorized}, \emph{approved}, \emph{privileged}, or
\emph{protected} therefore do not answer a basic causal question: which
actors can actually make a consequential action occur? This paper
provides an action-relative method for deriving which trust-domain
coalitions are sufficient to cause protected execution, defined as the
occurrence of a designated protected state transition. The framework
models components, powers, resources, boundaries, and alternative
realization structures; includes update, recovery, override,
disablement, and alternative invocation; and separates causal control
over execution from control over the authoritative account of an
operation. It derives inclusion-minimal sufficient coalitions and tests
whether claimed execution boundaries remain independent of designated
upstream domains. Cross-domain analytical cases illustrate the method.
In a split-control, release-intended, open-state, source-bounded
Havenlon protocol model, the ordinary witness requires five trust
domains, while certificate replacement yields a three-domain
inclusion-minimal known requirement set among source-enumerated protocol
witnesses; the Linux domain remains insufficient for the complete
transition. Deployed global non-bypassability and boundary-bound veto
coverage remain unresolved. The framework is a conceptual and analytical
tool. It does not certify implementations, establish deployment
security, guarantee complete discovery of hidden powers, or define
evidence-verification semantics.

\textbf{Keywords:} execution authority; trust domains; protected
execution; causal realization witnesses; lifecycle security; security
architecture

\section{1. Introduction}\label{sec-1}

\protect\phantomsection\label{sec-1-1}{} High-risk automated systems
increasingly combine policy engines, network services, protected
hardware or software components, delegated credentials, lifecycle
management, recovery, and updates. A request may be approved in one
service, transformed into a command in another, admitted by a hardware
boundary, and executed only after additional local conditions hold.
Exceptional powers may later replace credentials, change code or policy,
disable a restriction, or invoke a different route. In such systems,
component names and workflow diagrams do not answer the central causal
question: \textbf{who can actually cause the protected state
transition?}

\protect\phantomsection\label{sec-1-2}{} Established areas answer
important but different questions. Access control determines permission;
reference-monitor and complete-mediation work characterize protected
mediation; capabilities and credentials convey exercisable authority;
separation of duty and thresholds distribute selected contributions;
trusted-computing mechanisms protect execution; verification evaluates
specified properties; and provenance, attestation, and auditing allocate
or protect accounts of activity. This paper does not treat those
approaches as mistaken or incomplete within their own aims. It asks a
system-level question that their local outputs do not settle by
themselves: after ordinary workflow powers and admitted lifecycle or
recovery powers are composed, which trust-domain coalitions can make the
designated action happen?

\protect\phantomsection\label{sec-1-3}{} The paper therefore asks, in
plain terms, \textbf{who can make the protected action happen?} More
formally: \emph{How should authority over high-risk automated execution
be decomposed across actors and trust domains so that the domains
capable of initiating execution attempts are explicit and no designated
upstream domain can unilaterally cause the protected state transition?}
Initiation is not treated as inherently unsafe. An upstream domain may
legitimately propose, authorize, approve, and issue a command while
remaining insufficient for execution because another domain controls an
indispensable contribution. The answer is relative to the protected
action, system model, threat model, environmental assumptions,
trust-domain assignment, and admitted witness family.

\protect\phantomsection\label{sec-1-4}{} We develop an integrated
authority-decomposition framework around that question.

\protect\phantomsection\label{sec-1-5}{} The paper makes four principal
contributions:

\begin{enumerate}
\def\labelenumi{\arabic{enumi}.}
\tightlist
\item
  \textbf{An action-relative vocabulary and typed model.} The framework
  separates proposal, authorization, approval, command, policy decision,
  policy update, veto, execution, and evidence powers, and relates
  components, domains, powers, resources, boundaries, protected
  transitions, realization witnesses, and epistemic objects.
\item
  \textbf{A witness-to-coalition derivation.} Support-selected
  realization witnesses are projected onto controlling trust domains to
  derive execution authority and inclusion-minimal sufficient
  coalitions. Admitted update, recovery, override, disablement, and
  alternative-invocation powers are included through reconfiguration
  composition.
\item
  \textbf{Boundary, independence, and concentration diagnostics.} The
  framework distinguishes structural traversal from resistance to
  unilateral alteration, satisfaction, disablement, and bypass;
  separates causal execution independence from evidentiary independence;
  and derives recurring dangerous authority patterns from those
  distinctions.
\item
  \textbf{Cross-domain analytical demonstrations and a bounded
  implementation-informed instantiation.} Controlled cases test whether
  the outputs change when authority assignments change, while the
  source-grounded Havenlon analysis records what the identified protocol
  sources establish, conditionally support, or leave unresolved.
\end{enumerate}

Together, these contributions form an integrated authority-centered
organization and derivation. The paper does not claim a new
access-control system, security mechanism, cryptographic primitive,
graph primitive, provenance mechanism, or certification framework.

\protect\phantomsection\label{sec-1-6}{} Section 2 positions the
framework relative to the established mechanisms and models it uses.
Sections 3 and 4 define protected execution, execution authority, and
the authority taxonomy. Section 5 supplies the typed
witness-and-coalition model; Section 6 derives dangerous authority
patterns; and Section 7 states the causal and evidentiary independence
conditions. Section 8 presents controlled cross-domain contrasts.
Section 9 applies the framework to the source-bounded Havenlon protocol
model. Section 10 discusses interpretation and limitations, followed by
the conclusion. An extended technical supplement is distributed with
this preprint as ancillary material. It retains complete derivations,
full tables, secondary cases, and extended source-grounding material; no
principal definition, method, result, or qualification depends on it.

\section{2. Conceptual Foundations and Positioning}\label{sec-2}

The framework is integrative. It uses established mechanisms and
formalisms as inputs, then derives action-relative trust-domain
coalitions and boundary properties that no single local result settles
by itself.

\protect\phantomsection\label{sec-2-1}{} \textbf{Authorization and
access control} supply subjects, roles, permissions, sessions, and
separation-of-duty constraints \citep[§§2.1--2.3, 3, and
5]{sandhu-ferraiolo-kuhn2000}. The present question is causal rather
than normative: a permitted request may be refused or may still depend
on another domain, while an unauthorized route may nevertheless realize
the transition.

\protect\phantomsection\label{sec-2-2}{} \textbf{Reference monitors and
complete mediation} supply mediation, tamper resistance, and complete
checking
\citetext{\citealp[§§3.4--3.6]{anderson1972}; \citealp[§III.A]{saltzer-schroeder1975}}.
This paper separately asks whether every admitted witness traverses the
claimed boundary and who can alter its rule, satisfy its accepted
inputs, disable it, or invoke an alternative route.

\protect\phantomsection\label{sec-2-3}{} \textbf{Usage and lifecycle
control} show that decisions and mutable attributes may change before,
during, or after use \citep[§§3--4.2]{park-sandhu2004}. The framework
represents that insight by composing admitted update, recovery,
override, disablement, and other reconfiguration steps with later
realization witnesses.

\protect\phantomsection\label{sec-2-4}{} \textbf{Capabilities,
credentials, privileges, and delegation} represent exercisable rights
and scoped authorization artifacts
\citetext{\citealp[p.~6]{dennis-van-horn1966}; \citealp[§§1, 1.1--1.4,
3.3, and 7]{rfc6749}}. Here they become typed witness inputs; the
additional inquiry is whether they are merely prerequisites or
sufficient support, and who controls equivalent minting, replacement, or
recovery routes.

\protect\phantomsection\label{sec-2-5}{} \textbf{Separation of duty and
threshold mechanisms} distribute selected contributions
\citetext{\citealp[pp.~186--187]{clark-wilson1987}; \citealp[§5]{sandhu-ferraiolo-kuhn2000}; \citealp[pp.~612--613]{shamir1979}}.
Trust-domain projection then tests whether nominally distinct principals
remain independent under shared administration, recovery, key ownership,
or compromise.

\protect\phantomsection\label{sec-2-6}{} \textbf{Trusted computing bases
and lifecycle security} identify enforcement mechanisms and their
update, detection, and recovery roots \citetext{\citealp[§§3.2.3.2.3,
4.1.3.2.3, 6.1--6.3, and Glossary]{tcsec1985}; \citealp[§§3.1, 3.3--3.6,
and 4.1--4.4]{nist-sp800-193}}. The framework asks which lifecycle
controllers can change the witness inventory, boundary semantics, or
coalition requirements.

\protect\phantomsection\label{sec-2-7}{} \textbf{Attack, privilege,
AND/OR, hypergraph, and fault models} represent alternative routes,
conjunctive prerequisites, privilege expansion, and minimal sufficient
combinations \citetext{\citealp[executive summary and
§§2--3]{phillips-swiler1998}; \citealp[§§2--4]{dacier-deswarte1994}; \citealp{schneier1999}; \citealp[§§2--5]{gallo-et-al1993}; \citealp[chs.~IV
and VII]{vesely-et-al1981}}. This paper claims no new graph primitive;
it binds a typed, support-selected derivation to one protected
transition and projects its selected support to controlling domains.

\protect\phantomsection\label{sec-2-8}{} \textbf{HSMs, cryptographic
modules, enclaves, and interlocks} can supply concrete execution,
refusal, or local-condition contributions \citetext{\citealp[abstract
and §3.2]{fips140-3}; \citealp[§§1 and
3--5]{pkcs11-v3.1}; \citealp[pp.~1--5]{mckeen-et-al2013}; \citealp{hse-control-systems}}.
Their presence does not by itself establish that the contribution is
independently controlled or that lifecycle and alternative-invocation
paths remain outside the constrained coalition.

\protect\phantomsection\label{sec-2-9}{} \textbf{Provenance, logging,
attestation, and audit} define roles and protections for accounts of
activity \citetext{\citealp[§§1, 2.1, and
5.3]{w3c-prov-dm}; \citealp[abstract and
§§2--4]{schneier-kelsey1999}; \citealp[abstract and §§1 and
6--8]{rfc5848}; \citealp[§§3--4.2]{rfc9334}}. This paper allocates
control over evidence production, attestation, preservation,
suppression, and official designation. It does not define evidence
validity, trace completeness, outcome consistency, verifier correctness,
relation validity, or conformance.

\protect\phantomsection\label{sec-2-closest-work}{} The closest
foundations combine causal-authority analysis, attack or fault graph
reasoning, trust management, and lifecycle control
\citep{murray-lowe2007, sheyner-et-al2002, kordy-et-al2014, li-winsborough-mitchell2003, nist-sp800-193}.
The contribution is the common typed pipeline that connects these
foundations to protected-action-relative witnesses, trust-domain
coalitions, boundary diagnostics, and evidentiary-authority allocation.

\protect\phantomsection\label{tbl-2-001}{}

\emph{Table 2.1. Established support and the framework's integrated
output.}

{\def\LTcaptype{none} 
\begin{longtable}[]{@{}
  >{\raggedright\arraybackslash}p{(\linewidth - 2\tabcolsep) * \real{0.5000}}
  >{\raggedright\arraybackslash}p{(\linewidth - 2\tabcolsep) * \real{0.5000}}@{}}
\toprule\noalign{}
\begin{minipage}[b]{\linewidth}\raggedright
Established support
\end{minipage} & \begin{minipage}[b]{\linewidth}\raggedright
Integration in this paper
\end{minipage} \\
\midrule\noalign{}
\endhead
\bottomrule\noalign{}
\endlastfoot
Authorization, capabilities, and threshold control & Typed,
support-selected contributions projected to threat-model-relative trust
domains. \\
Attack, fault, and hypergraph structures & Grounded AND/OR/threshold
witnesses for one concrete protected transition. \\
Reference monitors and lifecycle control & Witness-relative traversal
plus separate alteration, satisfaction, disablement, and bypass
tests. \\
Provenance, logging, and attestation & A separate allocation of
evidentiary control without importing verification semantics. \\
\end{longtable}
}

\protect\phantomsection\label{sec-2-10}{} The residual question is
therefore compositional: which ordinary or reconfiguration-enabled
witnesses can realize the transition, which domain coalitions can supply
them, whether a claimed-final boundary constrains every witness, and
whether the execution coalition also controls the authoritative account.
The novelty claim is limited to organizing and deriving those outputs
around that action-relative question. The technical supplement provides
the extended area-by-area comparison.

\section{3. Protected Execution as an Authority-Governed State
Transition}\label{sec-3}

High-risk automated systems convert digital inputs into consequential
financial, operational, cryptographic, or physical state changes.
Authorization may precede those changes, but the event to be explained
is the occurrence of the designated transition, whether reached through
the intended workflow or another route.

This paper calls that occurrence \textbf{protected execution}.
\emph{Protected} marks the transition as an object of the threat model;
it does not imply authorization, correct mediation, or adequate
security, and it does not mean execution inside a trusted environment
\citep{globalplatform-tee-system-architecture2017, mckeen-et-al2013}.
The term is an analytical abstraction, not a new enforcement mechanism.
Reference monitors, usage-control systems, HSMs, interlocks, and
privileged services remain the concrete mechanisms
\citep{anderson1972, saltzer-schroeder1975, park-sandhu2004, nist-sp800-162, fips140-3, pkcs11-v3.1, hse-control-systems}.
The framework asks how the powers that initiate, permit, constrain,
reconfigure, perform, or attest the transition are distributed across
trust domains.

\subsection{3.1 From permission to state change}\label{sec-3-1}

Authorization asks whether a request is permitted under stated rules;
protected execution asks whether the designated transition occurred.
Permission may be necessary, but it is not occurrence.

Let \(r\) denote a request concerning an action \(a\), and let

\protect\phantomsection\label{eq-3-001}{}

\[
\mathsf{Authorize}(r,a,\gamma) \in \{\mathsf{permit},\mathsf{deny},\mathsf{conditional}\}
\]

represent the decision under context \(\gamma\). A permit result
establishes rule satisfaction at that point, not occurrence of the
action or control of its executing component.

The scope is relative to a threat model \(\Theta\). Let
\(\mathcal{A}_{P}(\Theta)\) contain the actions whose occurrence changes
state that \(\Theta\) classifies as consequential or security-sensitive.
This designation identifies what must be explained; it does not
establish that any mechanism protects it adequately.

For each \(a \in \mathcal{A}_{P}(\Theta)\), let

\protect\phantomsection\label{eq-3-002}{}

\[
\tau_a : s^{-}_{a} \rightarrow s^{+}_{a}
\]

denote its designated state transition, where \(s^{-}_{a}\) and
\(s^{+}_{a}\) are the action-relevant projections of system state before
and after the transition. For a system run \(\rho\),

\protect\phantomsection\label{eq-3-003}{}

\[
\mathsf{PE}(a,\rho) \iff
a \in \mathcal{A}_{P}(\Theta)
\land
\operatorname{Occurs}(\tau_a,\rho).
\]

Protected execution is therefore the occurrence of \(\tau_a\), not the
surrounding workflow's completion status. Later reversal does not erase
the fact that the transition occurred.

This distinction separates at least five analytically distinct stages,
conditions, or questions that are often compressed in system
descriptions:

\begin{enumerate}
\def\labelenumi{\arabic{enumi}.}
\tightlist
\item
  A request is \textbf{proposed}.
\item
  The request is \textbf{authorized or approved}.
\item
  An actionable \textbf{command} is issued toward an execution boundary.
\item
  The command produces an \textbf{execution attempt} at that boundary.
\item
  The designated protected state transition \textbf{occurs or does not
  occur}.
\end{enumerate}

The stages may involve different actors or collapse into one process;
either allocation is recorded because it changes the effect of
compromise.

An approved operation can therefore be refused without contradiction.
Refusal prevents \(\tau_a\) through the mediated path governed by that
boundary; it does not establish non-occurrence elsewhere in the run. It
shows only that upstream approval was insufficient through the refused
path.

\subsection{3.2 Execution attempts and protected
execution}\label{sec-3-2}

Dispatch, API invocation, or queue insertion is not protected execution
unless \(\tau_a\) occurs.

Formal claims use a \textbf{causal realization witness}: a possibly
non-linear derivation containing every selected AND, OR, threshold,
guard, resource, and reconfiguration condition needed for \(\tau_a\).
\emph{Path} refers only to an informal linear route.
\hyperref[sec-5]{Section 5} supplies the complete semantics.

A \textbf{candidate execution boundary} can accept or refuse an
actionable command before \(\tau_a\). A \textbf{claimed final execution
boundary} is a candidate that the design claims every feasible witness
traverses. It is \textbf{non-bypassable} only if at least one feasible
witness exists and every feasible witness traverses it.
Non-bypassability is therefore a non-vacuous derived property, not part
of the definition of a boundary or attempt
\citep{anderson1972, saltzer-schroeder1975}.

An \textbf{execution attempt} occurs when an actionable command reaches
a candidate execution boundary and requests that an action
\(a \in \mathcal{A}_{P}(\Theta)\) occur. If \(c\) is a command, \(B\) is
a candidate boundary, and \(\rho\) is a system run, then

\protect\phantomsection\label{eq-3-004}{}

\[
\mathsf{Attempt}(c,a,B,\rho) \iff
\mathsf{Reaches}(c,B,\rho)
\land
\mathsf{Requests}(c,a).
\]

The definition assumes neither non-bypassability nor correct
configuration. An attempt may be refused without protected execution,
and protected execution may occur without an attempt at the claimed
boundary.

This yields three analytically distinct cases:

\begin{itemize}
\tightlist
\item
  \textbf{Correctly mediated protected execution} occurs when the
  realized causal witness for \(\tau_a\) traverses the claimed final
  boundary and satisfies the mediation conditions specified for that
  boundary.
\item
  \textbf{Bypass execution} occurs, relative to a claimed final boundary
  \(B_a^{*}\), when a realized causal witness for \(\tau_a\) does not
  traverse \(B_a^{*}\).
\item
  \textbf{Improperly mediated protected execution} occurs when the
  realized causal witness traverses the claimed boundary but the
  transition is produced after that boundary has been compromised,
  disabled, misconfigured, or made to accept inputs outside its
  specified conditions.
\end{itemize}

All three are protected execution because the designated transition
occurred. The qualifiers describe how it occurred. In particular,

\protect\phantomsection\label{eq-3-005}{}

\[
\operatorname{BypassExec}(a,B_a^{*},\rho) \iff
\mathsf{PE}(a,\rho)
\land
\exists w\,
\bigl(
\operatorname{RealizedWitness}_{M,\Theta,\Gamma}(w,a,\rho)
\land
\neg\operatorname{Traverses}(w,B_a^{*})
\bigr).
\]

The predicate is witness-relative, not run-wide: one run may contain
both a refused mediated attempt and a distinct realized witness that
avoids \(B_a^{*}\). \hyperref[sec-5]{Section 5} defines feasibility and
run instantiation.

The model is mechanism-neutral: reference monitors, policy engines, HSM
policies, interlocks, capabilities, and threshold arrangements may all
supply mediation
\citep{nist-sp800-162, fips140-3, pkcs11-v3.1, hse-control-systems, clark-wilson1987, sandhu-ferraiolo-kuhn2000, shamir1979}.
Physical separation alone does not establish independent authority.

For composite or non-atomic operations, the analyst should model
separately each protected sub-transition and its commit point. A
workflow failure does not imply that no protected state change occurred.

Alternative interfaces, maintenance and debug facilities, updates, and
privileged invocation paths belong in the witness inventory. If any
admitted route enables a witness for \(\tau_a\) that avoids \(B_a^{*}\),
the boundary is bypassable. Questions about evidence validity,
completeness, verification, or conformance remain outside this paper.

\subsection{3.3 Execution authority}\label{sec-3-3}

\textbf{Authority} is a domain-controlled power that changes which
outcomes are reachable; it does not imply legitimacy. \textbf{Execution
authority} is the power to cause a designated protected transition. It
is distinct from permission, credential possession, workflow approval,
and policy evaluation, although those powers may contribute to a
realization.

The definition concerns effective rather than nominal control. An
``executor'' may remain insufficient because another domain can refuse,
while an ``administrator'' may be sufficient through update, recovery,
disablement, or alternative invocation. Execution authority may
therefore be unilateral, jointly sufficient, or distributed across
alternative minimal coalitions
\citep{clark-wilson1987, sandhu-ferraiolo-kuhn2000, shamir1979}.

Fix a system configuration \(M\), a threat model \(\Theta\), a set of
relevant trust domains \(\mathcal{D}\), and environmental assumptions
\(\Gamma\). For \(a \in \mathcal{A}_{P}(\Theta)\) and a coalition
\(X \subseteq \mathcal{D}\), the predicate

\protect\phantomsection\label{eq-3-006}{}

\[
\mathsf{EA}_{M,\Theta,\Gamma}(X,a)
\]

holds when \(X\) can cause \(\tau_a\) without an indispensable decision
from \(\mathcal{D}\setminus X\). Its attributed powers include ordinary
workflow powers and every update, override, credential-recovery,
disablement, alternative-invocation, or other reconfiguration power
admitted by \(\Theta\). Bypass remains derived: it exists only if those
powers enable a witness that avoids \(B_a^{*}\). The predicate states
capability, not inevitability; \hyperref[sec-5]{Section 5} gives its
witness semantics.

The assumptions in \(\Gamma\) may hold routine availability fixed but
must not hide an independently controlled choice. An outside service
that can still accept or refuse supplies an authority contribution, not
mere environment.

A \textbf{minimal sufficient execution-authority set} is a coalition
that can cause the transition and contains no smaller sufficient
coalition:

\protect\phantomsection\label{eq-3-007}{}

\[
\operatorname{MinimalEA}_{M,\Theta,\Gamma}(X,a)
\iff
\mathsf{EA}_{M,\Theta,\Gamma}(X,a)
\land
\forall Y \subsetneq X,\;
\neg\mathsf{EA}_{M,\Theta,\Gamma}(Y,a).
\]

The derived question is which inclusion-minimal domain sets can cause
the transition. A singleton establishes unilateral capability; if every
minimal set includes an independently controlled boundary domain,
upstream command authority alone is insufficient.
\hyperref[sec-5]{Sections 5} and \hyperref[sec-7]{7} formalize those
conclusions.

\subsection{3.4 Trust domains and unilateral causation}\label{sec-3-4}

A \textbf{trust domain} is a threat-model-relative boundary within which
compromise, administration, update authority, key ownership, or
organizational control is treated as shared. It need not coincide with a
machine, process, company, or location. Separate components collapse
into one domain under common control; components in one device may
remain separate only when their code, keys, lifecycle roots, and
invocation paths are independently controlled.

Domain \(D\) can \textbf{unilaterally cause protected execution} when
\(\mathsf{EA}_{M,\Theta,\Gamma}(\{D\},a)\) holds. A second component
does not create causal separation if \(D\) can alter its rule,
manufacture all accepted inputs, disable it, or route around it.

\textbf{Evidence authority} is separately the power to produce, attest,
preserve, suppress, or officially designate an operation account. It
becomes a causal contribution to a later action only when the system
uses that designation as an input to the later protected transition.

Unilateral causation is not inherently unacceptable; the framework
exposes it for comparison with the action's consequences and threat
model. Component count is irrelevant when one domain controls the
supposedly distinct approvals, credentials, policies, or executors. The
unit of analysis is independent control over indispensable conditions.

\subsection{3.5 Protected execution as an authority-governed state
transition}\label{sec-3-5}

Protected execution is an \textbf{authority-governed state transition}
because domain-controlled powers determine who can make \(\tau_a\)
occur.

An expected mediated path can be summarized as

\protect\phantomsection\label{eq-3-008}{}

\[
\begin{aligned}
\mathsf{Proposed}(r)
&\xrightarrow{\text{authorization or approval}}
\mathsf{Permitted}(r)\\
&\xrightarrow{\text{command issuance}}
\mathsf{Attempt}(c,a,B_a^{*},\rho)\\
&\xrightarrow{\text{boundary decision}}
\mathsf{Decision}_{B_a^{*}}(c,\rho),\\[2mm]
\mathsf{Decision}_{B_a^{*}}(c,\rho)=\mathsf{accept}
&\longrightarrow
\begin{cases}
\mathsf{Realizes}(\pi_m,\tau_a,\rho),\\
\neg\mathsf{Realizes}(\pi_m,\tau_a,\rho),
\end{cases}\\[2mm]
\mathsf{Decision}_{B_a^{*}}(c,\rho)=\mathsf{refuse}
&\longrightarrow
\neg\mathsf{Realizes}(\pi_m,\tau_a,\rho),
\end{aligned}
\]

Here \(\pi_m\) is one linear mediated path, not the witness object used
for formal authority, bypass, non-bypassability, or veto claims.
Acceptance does not establish protected execution unless \(\tau_a\)
occurs. Refusal establishes only non-occurrence through the refused
path, not elsewhere in the run.

Refusal may still change audit, retry, recovery, or policy state, and a
distinct witness \(w_b\) may realize \(\tau_a\) without traversing
\(B_a^{*}\), even in the same run. A compromised boundary may also be
traversed without applying its specified conditions. Each resulting
protected sub-transition must be modeled separately.

The diagram is an analytical skeleton, not a mandatory protocol. The
analyst must distinguish the transition from preceding decisions,
identify the domains controlling indispensable steps, and enumerate
every admitted witness. Access control, complete mediation, HSM policy,
and interlocks retain their native roles; the additional question is
whether lifecycle or alternative powers restore upstream causal control.
Authority decomposition does not prove safety, implementation
correctness, witness completeness, or resistance to collusion. It
establishes only which modeled domain coalitions are sufficient or
insufficient for the designated transition.

\section{4. Authority Taxonomy}\label{sec-4}

\hyperref[sec-3]{Section 3} defined protected execution as the
occurrence of a threat-model-designated state transition and authority
as a domain-controlled power that changes which outcomes are reachable.
This section distinguishes nine core authority types: proposal,
authorization, approval, command, policy decision, policy update, veto,
execution, and evidence authority. The taxonomy is neither exhaustive
nor mutually exclusive. It exposes powers that descriptions such as
``authorized,'' ``approved,'' or ``allowed to execute'' often collapse,
without prescribing a universal workflow or treating distribution as
proof of security.

\subsection{4.1 How to read the taxonomy}\label{sec-4-1}

An authority type classifies a power, not an actor, component, role,
message, or stage. Classification follows what a trust domain can make
possible or impossible. The same type may be distributed across domains;
one domain may hold several types; and one power may receive several
classifications when it performs several functions. Assignments are
relative to the protected action \(a\), model \(M\), threat model
\(\Theta\), environmental assumptions \(\Gamma\), and any governed path.
A firmware-replacement power, for example, may reconfigure action \(a\)
while its exercise is itself a separate protected action \(u\).

The labels also have different formal roles. Proposal, authorization,
approval, command, policy-decision, policy-update, and veto authority
classify powers that contribute to or constrain realizations. Execution
authority is an outcome-relative sufficiency property under
\(\mathsf{EA}_{M,\Theta,\Gamma}\). Evidence authority ordinarily
concerns control over accounts of an operation rather than causal
control of the described transition. All classifications are
descriptive: a legitimate service and an attacker controlling the same
command path may both possess command authority under \(\Theta\),
despite different normative status.

\protect\phantomsection\label{tbl-4-001}{}

\emph{Table 4.1. Protected-action-relative authority taxonomy.}

{\def\LTcaptype{none} 
\begin{longtable}[]{@{}
  >{\raggedright\arraybackslash}p{(\linewidth - 6\tabcolsep) * \real{0.2500}}
  >{\raggedright\arraybackslash}p{(\linewidth - 6\tabcolsep) * \real{0.2500}}
  >{\raggedright\arraybackslash}p{(\linewidth - 6\tabcolsep) * \real{0.2500}}
  >{\raggedright\arraybackslash}p{(\linewidth - 6\tabcolsep) * \real{0.2500}}@{}}
\toprule\noalign{}
\begin{minipage}[b]{\linewidth}\raggedright
Authority type
\end{minipage} & \begin{minipage}[b]{\linewidth}\raggedright
Object controlled
\end{minipage} & \begin{minipage}[b]{\linewidth}\raggedright
Analytical significance
\end{minipage} & \begin{minipage}[b]{\linewidth}\raggedright
What it does not establish
\end{minipage} \\
\midrule\noalign{}
\endhead
\bottomrule\noalign{}
\endlastfoot
Proposal & Creation or submission of a request & Introduces an operation
for consideration & Permission, attempt, or execution \\
Authorization & Normative permission status under a rule or entitlement
& Determines permit, deny, or conditional status & Approval of a
concrete instance or execution \\
Approval & Affirmation of a concrete request or workflow instance &
Supplies affirmative consent or confirmation & Policy correctness or
execution \\
Command & Production of an actionable instruction & Directs an operation
toward an execution boundary & Arrival at the boundary or protected
execution \\
Policy decision & Currently applicable rules and contextual inputs at a
decision point & Produces a decision that may instantiate authorization,
contribute to veto, qualify approval, or remain advisory & Authority to
change the rules \\
Policy update & Modification of rules governing later decisions &
Changes which later decisions or paths are reachable & Exercise of the
protected transition by itself \\
Veto & Refusal of an attempt or blockage of a governed path & Prevents
\(\tau_a\) through that mediated path & Non-occurrence through every
other path \\
Execution & Outcome-relative causal sufficiency for the designated
protected transition & Establishes that a domain or coalition can cause
\(\tau_a\) under \(\mathsf{EA}_{M,\Theta,\Gamma}\) & Authorization,
legitimacy, correct mediation, or uniqueness of the sufficient
coalition \\
Evidence & Production, attestation, preservation, suppression, or
official designation of evidence & Affects which operation accounts are
created, endorsed, retained, withheld, or treated as authoritative &
Evidence validity, verification, or causal control of the same
transition \\
\end{longtable}
}

The categories are functional rather than chronological. They may recur,
overlap, or be absent, and a bypass may omit the nominal proposal,
authorization, or approval sequence. Analysis therefore covers every
admitted control path rather than mapping the labels mechanically onto
one expected trace.

\subsection{4.2 Initiating authority: proposal and
command}\label{sec-4-2}

\textbf{Proposal authority} is the ability to create or submit a request
concerning an action, including its target, parameters, timing, and
requested operation where the interface permits. It does not establish
permission, command issuance, arrival at a boundary, or execution.

\textbf{Command authority} is the ability to turn a request, decision,
or operation into an actionable instruction directed at an execution
boundary. Its object is the command, not the request's legitimacy.
Issuance becomes an execution attempt only when the command reaches a
candidate boundary and requests the protected action; it does not itself
establish protected execution.

The distinction separates expression of intent from production of an
executable instruction. The powers may belong to different domains or
collapse in a direct-call architecture. Command authority also covers
maintenance, retry, recovery, CI/CD, administrative, and debug routes
admitted by \(\Theta\). Whether any such route contributes to a feasible
realization witness is determined by the typed model, not by its status
as an exceptional path.

\subsection{4.3 Normative, affirmative, and contextual decision
authority}\label{sec-4-3}

\textbf{Authorization authority} determines the normative permission
status of an actor or request under a rule or entitlement: permit, deny,
or conditional. It may be request-specific, but derives from the rule
being applied. Authorization is not inherently affirmative and does not
cause \(\tau_a\).

\textbf{Approval authority} affirms a concrete request or workflow
instance. A threshold establishes only that the approval condition has
been satisfied; it does not establish later permission or execution.
Multiple principals do not imply multiple trust domains when they share
administration, recovery, update control, key ownership, or another
compromise boundary under \(\Theta\).

Authorization and approval may be exercised through one interface, but
remain analytically distinct because delegation, compromise, and update
can affect them differently. A domain may approve an instance without
controlling its governing entitlement, or control the entitlement
without affirming the instance.

\textbf{Policy decision authority} evaluates currently applicable rules
and contextual inputs at a decision point. Its output may instantiate
authorization, qualify approval, contribute to a veto, or remain
advisory, depending on its downstream effect. One decision may therefore
receive several classifications. A negative output is not automatically
a veto: a non-binding result has policy-decision authority only if it
can affect a later decision or realization, whereas veto authority
requires control sufficient to block the governed mediated path. Purely
informational output that no downstream function consumes is not
authority under the descriptive definition.

\subsection{4.4 Meta-authority: policy update and related
control}\label{sec-4-4}

\textbf{Policy-update authority} changes the rules used by later
authorization, approval, policy-decision, or veto functions, including
thresholds, accepted identities, targets, limits, and acceptance
conditions. It differs from deciding under the rules currently in force.

Firmware update, key replacement, credential recovery, boundary
disablement, override, and alternative invocation are broader
reconfiguration powers, not implicit subtypes of policy update.
\hyperref[sec-5]{Section 5} represents them explicitly. Bypass is
derived only when admitted powers enable a realization witness for
\(\tau_a\) that avoids \(B_a^{*}\); there is no primitive
\texttt{CAN\_BYPASS} relation. Neither policy update nor broader
meta-control necessarily establishes execution authority alone, but both
may change the sufficient coalitions and must therefore appear in the
causal analysis.

\subsection{4.5 Restrictive authority: veto}\label{sec-4-5}

\textbf{Veto authority} is the ability to refuse an attempt or prevent a
specified realization of \(\tau_a\) through a governed mediated path. It
is not the negative form of approval: prior approvals may remain valid
even when a later boundary blocks their attempted realization.

Veto is always path-relative. Blocking command issuance or refusing at a
candidate boundary prevents that governed realization, but not
necessarily a maintenance route, second command path, compromised
boundary, or other executor. Global prevention requires at least one
feasible witness, coverage of every feasible witness, and the relevant
independence conditions. Non-bypassability and independence are
therefore properties to be derived, not consequences of labeling a
denial, HSM refusal, interlock, or service rejection as a veto.

\subsection{4.6 Transition-realizing authority:
execution}\label{sec-4-6}

\textbf{Execution authority} is the outcome-relative property of a
trust-domain coalition sufficient to cause \(\tau_a\) under
\(\mathsf{EA}_{M,\Theta,\Gamma}\). Protected execution remains execution
even when unauthorized, incorrectly mediated, or reached through bypass.
The nominal executor may be insufficient without another domain's
indispensable decision, while an administrator with direct invocation,
replacement, or disablement powers may belong to a sufficient coalition.

A domain has \textbf{unilateral execution authority} when its singleton
coalition satisfies \(\mathsf{EA}\). A coalition is \textbf{jointly
sufficient} when it satisfies \(\mathsf{EA}\) while no member is
sufficient alone; inclusion-minimality is assessed separately. Several
alternative minimal sufficient coalitions may exist, and finding one
singleton does not exclude others. \hyperref[sec-5]{Section 5} derives
these coalitions from feasible witnesses rather than from component
names or the nominal workflow. The result identifies concentration and
dependency; it does not imply that either isolation or distribution is
safe.

\subsection{4.7 Epistemic authority: evidence}\label{sec-4-7}

\textbf{Evidence authority} controls production, attestation,
preservation, suppression, or official designation of evidence
associated with an operation. These powers govern which accounts are
created, endorsed, retained, withheld, or treated as authoritative; they
need not be co-located and do not ordinarily determine whether the
described transition occurs.

Evidence authority remains distinct from execution authority even when
one domain holds both. It contributes causally to a later action only
when the system treats an evidentiary designation as an input to that
later protected transition. The contribution then concerns the later
action. This separation permits analysis of self-attestation and
evidence concentration without defining evidence validity, trace
completeness, outcome consistency, verifier semantics, or conformance;
\hyperref[sec-7]{Section 7} treats evidentiary independence as an
accountability property separate from causal veto independence.

\subsection{4.8 Classification rules and analytical
limits}\label{sec-4-8}

Apply the taxonomy using six rules:

\begin{enumerate}
\def\labelenumi{\arabic{enumi}.}
\tightlist
\item
  \textbf{Classify effects, not labels.} Determine what a domain can
  make reachable or unreachable for \(a\) under \(M,\Theta,\Gamma\), and
  the relevant path.
\item
  \textbf{Permit overlap.} One power may instantiate several types, one
  domain may hold several types, and domain-specific subtypes may be
  added where analytically necessary.
\item
  \textbf{Represent exceptional powers explicitly.} Include admitted
  maintenance, recovery, debug, update, replacement, disablement,
  override, and alternative-invocation powers.
\item
  \textbf{Separate path functions from outcome sufficiency.} The
  taxonomy identifies powers; \(\mathsf{EA}_{M,\Theta,\Gamma}\) derives
  unilateral, jointly sufficient, and alternative minimal sufficient
  coalitions from the typed model and feasible-witness inventory.
\item
  \textbf{Keep restrictive claims path-relative.} A refusal becomes
  global prevention only with witness coverage, non-vacuity, and
  independence.
\item
  \textbf{Keep evidence authority epistemic unless it controls a later
  transition.} Authority over an account is not causal control of the
  operation described.
\end{enumerate}

The framework does not require all nine types, a fixed order, or
separate components. It requires analysts to distinguish the powers that
exist and expose their concentration and dependencies.

\section{5. Authority Graph Model}\label{sec-5}

This section turns the preceding vocabulary into a typed model for
deriving sufficient trust-domain coalitions, boundary traversal, and
veto coverage. ``Authority graph'' is convenient shorthand, but the
causal plane is a directed hypergraph rather than an actor-to-actor
graph. The model is mechanism-neutral and derives authority from
controlled powers and feasible realization witnesses, not component or
role names.

Here \(a\) is a \textbf{concrete protected-action instance}, such as one
specified payment, signature, or actuation request. Repeated operations
of the same type are distinct instances, each mapped to its own
transition. This convention binds the request, witness, occurrence, and
bypass claim to the same action.

\subsection{5.1 Why an ordinary authority graph is
insufficient}\label{sec-5-1}

Ordinary directed reachability cannot distinguish conjunctive
prerequisites, alternative derivations, or threshold subsets. Component
diagrams also exaggerate independence when several boxes share
administration, update, recovery, or another compromise boundary.
Finally, nominal workflow edges omit reconfiguration powers that may
create alternative realizations. The model therefore uses typed
hyperarcs, trust-domain projection, explicit power and resource control,
reconfiguration, and support-selected witnesses
\citep{gallo-et-al1993, schneier1999, phillips-swiler1998, vesely-et-al1981}.
It identifies sufficiency under declared assumptions rather than
simulating an implementation.

\subsection{5.2 Typed analytical objects and trust-domain normal
form}\label{sec-5-2}

For a system configuration \(M\), threat model \(\Theta\), environmental
assumptions \(\Gamma\), and concrete protected-action instance \(a\),
define the typed authority model

\protect\phantomsection\label{eq-5-001}{}

\[
\mathcal{G}_{M,\Theta,\Gamma,a}
=
\left\langle
\mathcal{C},
\mathcal{D},
\mathcal{P},
\mathcal{A}_{P},
\mathcal{T},
\mathcal{B}_{C},
\mathcal{B}_{F},
\mathcal{R},
\mathcal{E},
\delta,
\kappa,
\tau,
\operatorname{Class},
\mathcal{H}_{C},
\mathcal{H}_{E}
\right\rangle .
\]

Here \(\mathcal{C}\) contains analytical components; \(\mathcal{D}\),
trust domains; \(\mathcal{P}\), controlled powers;
\(\mathcal{A}_{P}(\Theta)\), protected-action instances;
\(\mathcal{T}\), designated transitions; \(\mathcal{B}_{C}\) and
\(\mathcal{B}_{F}\), action-indexed candidate and claimed-final boundary
sets; \(\mathcal{R}\), controlled resources and mutable state;
\(\mathcal{E}\), epistemic objects; and
\(\mathcal{H}_{C},\mathcal{H}_{E}\), the causal and epistemic typed
hypergraphs.

The protected-action mapping has the signature

\protect\phantomsection\label{eq-5-002}{}

\[
\tau:\mathcal{A}_{P}(\Theta)\rightarrow\mathcal{T},
\qquad
\tau(a)=\tau_a.
\]

Thus \(\tau_a\) denotes the transition for the same concrete instance in
every witness and run-level predicate.

Boundary status is also action-relative. Formally,

\protect\phantomsection\label{eq-5-003}{}

\[
\mathcal{B}_{C},\mathcal{B}_{F}:
\mathcal{A}_{P}(\Theta)\rightarrow 2^{\mathcal{C}},
\]

and

\protect\phantomsection\label{eq-5-004}{}

\[
\begin{aligned}
\operatorname{CandidateFor}_{M,\Theta}(B,a)
&\iff B\in\mathcal{B}_{C}(a),\\
\operatorname{ClaimedFinalFor}_{M,\Theta}(B,a)
&\iff B\in\mathcal{B}_{F}(a).
\end{aligned}
\]

A candidate boundary can accept or refuse an actionable command before
\(\tau_a\); a claimed-final boundary is asserted by the design to lie on
every feasible witness. Hence

\protect\phantomsection\label{eq-5-005}{}

\[
\forall a\in\mathcal{A}_{P}(\Theta),\qquad
\mathcal{B}_{F}(a)\subseteq\mathcal{B}_{C}(a),
\]

or, equivalently,

\protect\phantomsection\label{eq-5-006}{}

\[
\operatorname{ClaimedFinalFor}_{M,\Theta}(B,a)
\Rightarrow
\operatorname{CandidateFor}_{M,\Theta}(B,a).
\]

The designation records, but does not prove, an action-relative
architectural claim. \(B_a^{*}\) denotes a distinguished member of
\(\mathcal{B}_{F}(a)\), not a globally final boundary.

The component-to-domain assignment is

\protect\phantomsection\label{eq-5-007}{}

\[
\delta_{M,\Theta}:\mathcal{C}\rightarrow\mathcal{D}.
\]

The mapping is single-valued because \(\mathcal{C}\) is placed in
\textbf{trust-domain normal form}: independently controlled surfaces of
one physical component are represented as analytical subcomponents until
each has one domain assignment. This is analytical decomposition, not a
claim of physical separation.

Control of powers and mutable resources is recorded separately:

\protect\phantomsection\label{eq-5-008}{}

\[
\kappa_{M,\Theta}:
\mathcal{P}\cup\mathcal{R}
\rightarrow
2^{\mathcal{D}}\setminus\{\varnothing\}.
\]

The set \(\kappa(p)\) or \(\kappa(r)\) identifies participating
controllers; hyperarcs specify whether their control is conjunctive,
disjunctive, or threshold-based. Components mapped by \(\delta\) to the
same domain contribute only one independent coalition member.

\(\Gamma\) may fix non-discretionary operating conditions but may not
conceal an independently controlled choice. Any domain able to accept or
refuse an indispensable contribution must appear in the causal model.

Table 5.1 separates model-level feasibility or capability from run-level
instantiation and occurrence.

\protect\phantomsection\label{tbl-5-001}{}

\emph{Table 5.1. Core model-level, witness-level, and run-level
notation.}

{\def\LTcaptype{none} 
\begin{longtable}[]{@{}
  >{\raggedright\arraybackslash}p{(\linewidth - 4\tabcolsep) * \real{0.3333}}
  >{\raggedright\arraybackslash}p{(\linewidth - 4\tabcolsep) * \real{0.3333}}
  >{\raggedright\arraybackslash}p{(\linewidth - 4\tabcolsep) * \real{0.3333}}@{}}
\toprule\noalign{}
\begin{minipage}[b]{\linewidth}\raggedright
Notation
\end{minipage} & \begin{minipage}[b]{\linewidth}\raggedright
Level
\end{minipage} & \begin{minipage}[b]{\linewidth}\raggedright
Meaning
\end{minipage} \\
\midrule\noalign{}
\endhead
\bottomrule\noalign{}
\endlastfoot
\(\mathcal{G}_{M,\Theta,\Gamma,a}\) & Model & Typed authority model for
concrete protected-action instance \(a\). \\
\(\tau(a)=\tau_a\) & Model & Maps \(a\) to its designated protected
state transition. \\
\(\delta_{M,\Theta}\) & Model & Assigns each trust-domain-normal-form
component to one domain. \\
\(\kappa_{M,\Theta}\) & Model & Identifies controller domains for powers
and mutable resources. \\
\(\operatorname{CandidateFor}_{M,\Theta}(B,a)\) & Model & \(B\) can
mediate an attempt concerning \(a\). \\
\(\operatorname{ClaimedFinalFor}_{M,\Theta}(B,a)\) & Model & The design
claims \(B\) lies on every feasible witness for \(a\). \\
\(\operatorname{Class}_{M,\Theta,\Gamma}(p,a,q)\) & Model & Core
authority labels assigned to power \(p\) relative to \(a\) and optional
governed causal substructure \(q\). \\
\(\begin{gathered}
\eta_{\mathrm{produce}},\eta_{\mathrm{attest}},\eta_{\mathrm{preserve}},\\
\eta_{\mathrm{suppress}},\eta_{\mathrm{official}}
\end{gathered}\)
& Model & Separate domain-control projections for the five evidence
powers. \\
\(\operatorname{FeasibleWitness}_{M,\Theta,\Gamma}(w,a)\) & Model &
\(w\) is a type-correct, support-selected feasible derivation of
\(\tau_a\). \\
\(\operatorname{HasFeasibleWitness}_{M,\Theta,\Gamma}(a)\) & Model & At
least one feasible witness for \(a\) exists. \\
\(\operatorname{Req}_{M,\Theta,\Gamma}(w)\) & Model & Independent trust
domains required by the selected support of \(w\). \\
\(\operatorname{CoalitionRealizable}_{M,\Theta,\Gamma}(w\mid X)\) &
Model & \(\operatorname{Req}(w)\subseteq X\). \\
\(\mathsf{EA}_{M,\Theta,\Gamma}(X,a)\) & Model & Some feasible witness
for \(a\) has domain requirements contained in \(X\). \\
\(\operatorname{MinimalEA}_{M,\Theta,\Gamma}(X,a)\) & Model & \(X\) has
execution authority and no proper subset does. \\
\(\operatorname{Traverses}(w,B)\) & Witness structure & Mediation by
\(B\) belongs to the selected causal support of \(w\). \\
\(\operatorname{Covers}(v,w)\) & Model / counterfactual & Exercising
veto \(v\) invalidates witness \(w\)'s selected support. \\
\(\operatorname{NonBypassable}_{M,\Theta,\Gamma}(B,a)\) & Model & A
claimed-final boundary is traversed by every member of a nonempty
feasible-witness set. \\
\(\operatorname{GlobalVetoCoverage}_{M,\Theta,\Gamma}(v,a)\) & Model &
Veto \(v\) covers every member of a nonempty feasible-witness set. \\
\(\operatorname{Instantiates}(\rho,\operatorname{supp}(w))\) & Run & Run
\(\rho\) instantiates the selected powers, facts, changes, and
dependencies in \(w\). \\
\(\operatorname{CausallyCulminates}(w,\tau_a,\rho)\) & Run & The
instantiated selected support of \(w\) causally culminates in \(\tau_a\)
in \(\rho\). \\
\(\operatorname{Occurs}(\tau_a,\rho)\) & Run & The transition for the
same concrete action instance \(a\) occurs in \(\rho\). \\
\(\mathsf{PE}(a,\rho)\) & Run & The concrete protected-action instance
is in scope and its designated transition occurs. \\
\(\operatorname{RealizedWitness}_{M,\Theta,\Gamma}(w,a,\rho)\) & Run & A
feasible witness is instantiated in \(\rho\) and causally culminates in
\(\tau_a\). \\
\(\operatorname{BypassExec}_{M,\Theta,\Gamma}(a,B,\rho)\) & Run & A
realized witness for \(a\) in \(\rho\) avoids boundary \(B\) claimed
final for \(a\). \\
\end{longtable}
}

\subsection{5.3 Power classification and primitive
relations}\label{sec-5-3}

The taxonomy in \hyperref[sec-4]{Section 4} classifies powers rather
than components. Let

\protect\phantomsection\label{eq-5-009}{}

\[
\mathcal{L}_{\mathrm{core}}
=
\{
\textsf{proposal},
\textsf{authorization},
\textsf{approval},
\textsf{command},
\textsf{policy\mbox{-}decision},
\textsf{policy\mbox{-}update},
\textsf{veto},
\textsf{evidence}
\}.
\]

The classification function is

\protect\phantomsection\label{eq-5-010}{}

\[
\operatorname{Class}_{M,\Theta,\Gamma}:
\mathcal{P}
\times
\mathcal{A}_{P}(\Theta)
\times
\bigl(\operatorname{Sub}(\mathcal{H}_{C})\cup\{\bot\}\bigr)
\rightarrow
2^{\mathcal{L}_{\mathrm{core}}}.
\]

The optional \(q\) identifies a governed causal substructure; use
\(q=\bot\) when none is required. Classifications may overlap. Execution
is excluded from \(\mathcal{L}_{\mathrm{core}}\) because it is derived
for coalitions through \(\mathsf{EA}_{M,\Theta,\Gamma}\), not assigned
to an isolated power.

Table 5.2 states the required primitive relation families;
domain-specific models may refine them where causal support differs.

\protect\phantomsection\label{tbl-5-002}{}

\emph{Table 5.2. Primitive power and relation families used in the
causal model.}

{\def\LTcaptype{none} 
\begin{longtable}[]{@{}
  >{\raggedright\arraybackslash}p{(\linewidth - 4\tabcolsep) * \real{0.3333}}
  >{\raggedright\arraybackslash}p{(\linewidth - 4\tabcolsep) * \real{0.3333}}
  >{\raggedright\arraybackslash}p{(\linewidth - 4\tabcolsep) * \real{0.3333}}@{}}
\toprule\noalign{}
\begin{minipage}[b]{\linewidth}\raggedright
Primitive power or relation family
\end{minipage} & \begin{minipage}[b]{\linewidth}\raggedright
Core classification when applicable
\end{minipage} & \begin{minipage}[b]{\linewidth}\raggedright
Modeled effect
\end{minipage} \\
\midrule\noalign{}
\endhead
\bottomrule\noalign{}
\endlastfoot
Request creation or submission & proposal & Introduces a request
concerning \(a\). \\
Determination of permit, deny, or conditional status & authorization;
possibly policy-decision & Establishes a normative status consumed by a
later structure. \\
Concrete affirmation or threshold contribution & approval & Supplies
instance-specific approval support. \\
Contextual rule evaluation & policy-decision; possibly authorization or
veto & Supplies a decision consumed by a later causal structure. \\
Command construction, issuance, delivery, or alternative invocation &
command & Creates or transports an actionable instruction. \\
Boundary acceptance or non-refusal & policy-decision or another
domain-specific label & Supplies selected support for a mediated
realization. \\
Boundary refusal & veto; possibly policy-decision & Invalidates the
governed mediated structure. \\
\texttt{CAN\_PERFORM\_TRANSITION} & no execution label assigned by
\(\operatorname{Class}\) & Performs \(\tau_a\) when its modeled
prerequisites hold. \\
Policy modification & policy-update & Changes rules used by later
decisions. \\
Firmware replacement, update-root control, credential or key replacement
and recovery, disablement, override & a core label only when its actual
effect warrants one; otherwise an explicit domain-specific
reconfiguration power & Changes configurations, guards, resources, or
available hyperarcs. \\
\end{longtable}
}

The final row is not collapsed into policy update because different
reconfiguration powers can produce different successor configurations
and witnesses. Likewise, \texttt{CAN\_PERFORM\_TRANSITION} states only
that a power can perform \(\tau_a\) when its prerequisites hold;
coalition sufficiency requires a complete feasible witness.

The epistemic primitive vocabulary records five powers separately:

\protect\phantomsection\label{eq-5-011}{}

\[
\begin{gathered}
\texttt{CAN\_PRODUCE\_EVIDENCE},\quad
\texttt{CAN\_ATTEST\_EVIDENCE},\quad
\texttt{CAN\_PRESERVE\_EVIDENCE},\\
\texttt{CAN\_SUPPRESS\_EVIDENCE},\quad
\texttt{CAN\_DESIGNATE\_OFFICIAL}.
\end{gathered}
\]

For each evidence object \(e\in\mathcal{E}\), the corresponding control
projections are

\protect\phantomsection\label{eq-5-012}{}

\[
\eta_{\mathrm{produce}},
\eta_{\mathrm{attest}},
\eta_{\mathrm{preserve}},
\eta_{\mathrm{suppress}},
\eta_{\mathrm{official}}
:
\mathcal{E}\rightarrow 2^{\mathcal{D}}.
\]

Each \(\eta\)-mapping projects controllers of one evidence function from
\(\mathcal{H}_{E}\) and \(\kappa\). It identifies control but does not
define evidence structure or validity. An official designation
contributes causally only when consumed by a later protected action, and
then only to that later action. Bypass is likewise non-primitive: it
exists only when admitted powers support a realized witness that avoids
a claimed-final boundary.

\subsection{5.4 Causal and epistemic hypergraphs}\label{sec-5-4}

\(\mathcal{H}_{C}\) contains typed power exercises, commands, decisions,
conditions, mediation, configurations, and transitions; its hyperarcs
derive causal facts from selected prerequisite sets. \(\mathcal{H}_{E}\)
contains epistemic objects and exercises of the five evidence powers,
with \(\eta\) projecting their controllers. This plane records control
over accounts, not their structure, relation validity, completeness,
consistency, verification, or conformance. An epistemic relation becomes
causal only through an explicit consumption relation from an official
designation to a later protected action.

\begin{figure}[p]
\centering
\protect\phantomsection\label{fig-5-1}
\includegraphics[width=\textwidth,height=0.76\textheight,keepaspectratio]{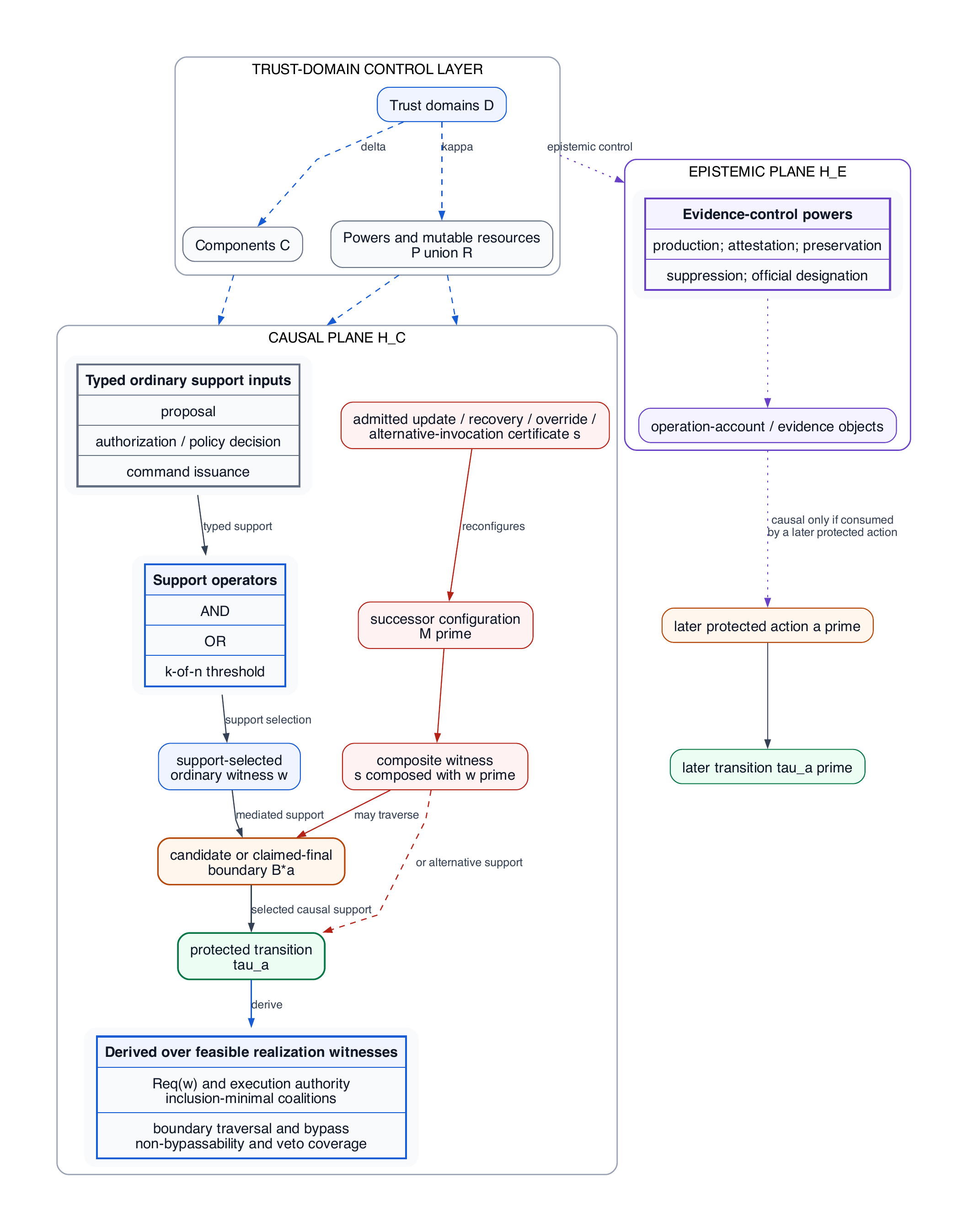}
\caption*{\textit{Figure 5.1. Layered authority model. AND, OR,
threshold, ordinary-witness, reconfiguration-witness, and
causal/epistemic structures remain explicit; the diagram does not assert
that every displayed alternative is present in every system.}}
\end{figure}

Figure \hyperref[fig-5-1]{5.1} summarizes the layered model.

The planes are semantically, not necessarily physically, separate: one
domain may control nodes in both.

\subsection{5.5 AND, OR, and threshold semantics}\label{sec-5-5}

The causal hypergraph uses conjunctive, disjunctive, and threshold
support.

A conjunctive hyperarc

\protect\phantomsection\label{eq-5-013}{}

\[
e=(S,h,\phi,\ell)
\]

derives \(h\) only when every typed prerequisite in \(S\) is satisfied,
guard \(\phi\) holds, and relation type \(\ell\) permits the derivation.

Disjunction uses multiple incoming hyperarcs; a witness selects one
supporting alternative.

A threshold hyperarc

\protect\phantomsection\label{eq-5-014}{}

\[
e=(S,k,h,\phi,\ell)
\]

derives \(h\) when a satisfying \(k\)-of-\(n\) subset of \(S\) is
present and the guard holds. A witness selects an inclusion-minimal
satisfying subset---exactly \(k\) prerequisites for an unweighted rule.

Support selection yields a definite causal certificate and excludes
unused branches from traversal, coalition, and veto-coverage predicates.

Thresholds are first evaluated at the system's principal or component
level, then projected through \(\delta_{M,\Theta}\). Three selected
approvers in one trust domain may satisfy the implementation threshold
while contributing only one independent domain to the authority
coalition
\citep{clark-wilson1987, sandhu-ferraiolo-kuhn2000, shamir1979}.

\subsection{5.6 Reconfiguration sequences}\label{sec-5-6}

A feasible realization from an initial configuration may include a
finite reconfiguration prefix:

\protect\phantomsection\label{eq-5-015}{}

\[
M_0
\xrightarrow{p_1}
M_1
\xrightarrow{p_2}
\cdots
\xrightarrow{p_j}
M_j
\xRightarrow{w}
\tau_a.
\]

Each \(p_i\) is admitted by \(\Theta\) and may change a guard, resource,
credential, boundary, interface, control assignment, or available
hyperarc. The suffix is evaluated in \(M_j\), while its coalition
includes controllers of every selected prefix step. Policy update,
firmware replacement, recovery, disablement, and override remain
distinct whenever they produce different successor configurations,
requirements, or traversal results. \(\Theta\) determines which
sequences are admitted; consequential reconfiguration may itself be
analyzed as another protected action. Final-boundary claims quantify
over all admitted witnesses from \(M\), including those enabled by such
prefixes.

\subsection{5.7 Causal realization witnesses}\label{sec-5-7}

A \textbf{causal realization witness} \(w\) is a finite,
support-selected certificate consisting of an optional reconfiguration
prefix and a type-correct derivation subhypergraph terminating in
\(\tau_a\). Its selected steps, powers, facts, nodes, and hyperarcs form
\(\operatorname{supp}(w)\).

The support must satisfy eight conditions:

\begin{enumerate}
\def\labelenumi{\arabic{enumi}.}
\tightlist
\item
  Its terminal conclusion is \(\tau_a\) for the same action instance.
\item
  Every AND prerequisite is retained.
\item
  Each OR step selects one supporting alternative.
\item
  Each threshold step selects one inclusion-minimal satisfying subset.
\item
  Selected guards and resource conditions hold under the modeled
  configuration and \(\Gamma\).
\item
  A strict order \(\prec_w\) places every selected prerequisite before
  its head.
\item
  Every leaf is grounded in the initial configuration, \(\Gamma\), a
  satisfied condition, or a controlled power exercise.
\item
  Every included element contributes to \(\tau_a\); unused alternatives
  and unrelated events are excluded.
\end{enumerate}

Well-foundedness and grounding exclude unsupported cycles. Traversal,
coverage, and coalition attribution are evaluated only over
\(\operatorname{supp}(w)\).

The model-level predicate

\protect\phantomsection\label{eq-5-016}{}

\[
\operatorname{FeasibleWitness}_{M,\Theta,\Gamma}(w,a)
\]

holds when \(a\in\mathcal{A}_{P}(\Theta)\) and \(w\) is admitted from
\(M\) under \(\Theta,\Gamma\), with jointly satisfiable controlled
prerequisites. Feasibility neither asserts run-level instantiation nor
selects a smallest coalition.

Nonempty realizability is explicit:

\protect\phantomsection\label{eq-5-017}{}

\[
\begin{aligned}
\operatorname{HasFeasibleWitness}_{M,\Theta,\Gamma}(a)
\iff\;&
a\in\mathcal{A}_{P}(\Theta)\\
&\land
\exists w\;
\operatorname{FeasibleWitness}_{M,\Theta,\Gamma}(w,a).
\end{aligned}
\]

At run level,

\protect\phantomsection\label{eq-5-018}{}

\[
\operatorname{Instantiates}(\rho,\operatorname{supp}(w))
\]

holds when \(\rho\) contains the selected exercises, changes, facts,
guards, and dependencies in \(\operatorname{supp}(w)\), all for the same
action instance. Other run events are excluded.

Mere co-occurrence is insufficient. The predicate

\protect\phantomsection\label{eq-5-019}{}

\[
\operatorname{CausallyCulminates}(w,\tau_a,\rho)
\]

requires the instantiated derivation to culminate causally in
\(\tau_a\), and entails \(\operatorname{Occurs}(\tau_a,\rho)\). Thus

\protect\phantomsection\label{eq-5-020}{}

\[
\begin{aligned}
\operatorname{RealizedWitness}_{M,\Theta,\Gamma}(w,a,\rho)
\iff\;&
\operatorname{FeasibleWitness}_{M,\Theta,\Gamma}(w,a)\\
&\land
\operatorname{Instantiates}(\rho,\operatorname{supp}(w))\\
&\land
\operatorname{CausallyCulminates}(w,\tau_a,\rho).
\end{aligned}
\]

The last conjunct excludes attempted derivations and unrelated
occurrences.

Coalition attribution remains model-level. For each feasible witness,
define

\protect\phantomsection\label{eq-5-021}{}

\[
\operatorname{Req}_{M,\Theta,\Gamma}(w)\subseteq\mathcal{D}
\]

as the independent domains required by \(\operatorname{supp}(w)\).
Construction projects selected components through \(\delta\), powers and
resources through \(\kappa\), follows selected controller alternatives
or thresholds, collapses duplicate domains, excludes only
non-discretionary facts in \(\Gamma\), and includes every selected
reconfiguration step.

For \(X\subseteq\mathcal{D}\), coalition realizability is

\protect\phantomsection\label{eq-5-022}{}

\[
\operatorname{CoalitionRealizable}_{M,\Theta,\Gamma}(w\mid X)
\iff
\operatorname{Req}_{M,\Theta,\Gamma}(w)\subseteq X.
\]

Unused alternatives and duplicate same-domain components add no
requirements; extra members of \(X\) need not act.

Finally,

\protect\phantomsection\label{eq-5-023}{}

\[
\operatorname{Traverses}(w,B)
\]

holds only when mediation by \(B\) belongs to the selected support
deriving \(\tau_a\). Appearance elsewhere, on an unused branch, or as a
disablement target is insufficient. One run may contain a refused route
through \(B\) and a distinct realized witness that avoids it.

\subsection{5.8 Execution-authority and minimal-coalition
derivation}\label{sec-5-8}

Execution authority holds when a coalition contains the requirements of
at least one feasible witness:

\protect\phantomsection\label{eq-5-024}{}

\[
\begin{aligned}
\mathsf{EA}_{M,\Theta,\Gamma}(X,a)
\iff\;&
\exists w\;
\bigl(
\operatorname{FeasibleWitness}_{M,\Theta,\Gamma}(w,a)
\land
\operatorname{Req}_{M,\Theta,\Gamma}(w)\subseteq X
\bigr).
\end{aligned}
\]

This is a monotone capability claim, not an assertion that a run
occurred.

A domain has unilateral execution authority when its singleton coalition
is sufficient:

\protect\phantomsection\label{eq-5-025}{}

\[
\operatorname{UnilateralEA}_{M,\Theta,\Gamma}(d,a)
\iff
\mathsf{EA}_{M,\Theta,\Gamma}(\{d\},a).
\]

A coalition is jointly sufficient when it is sufficient, contains more
than one domain, and no individual member is sufficient alone:

\protect\phantomsection\label{eq-5-026}{}

\[
\begin{aligned}
\operatorname{JointlySufficient}_{M,\Theta,\Gamma}(X,a)
\iff\;&
\mathsf{EA}_{M,\Theta,\Gamma}(X,a)
\land |X|>1\\
&\land
\forall d\in X,\;
\neg\mathsf{EA}_{M,\Theta,\Gamma}(\{d\},a).
\end{aligned}
\]

Joint sufficiency does not imply inclusion-minimality, which requires:

\protect\phantomsection\label{eq-5-027}{}

\[
\begin{aligned}
\operatorname{MinimalEA}_{M,\Theta,\Gamma}(X,a)
\iff\;&
\mathsf{EA}_{M,\Theta,\Gamma}(X,a)\\
&\land
\forall Y\subsetneq X,\;
\neg\mathsf{EA}_{M,\Theta,\Gamma}(Y,a).
\end{aligned}
\]

Equivalently, all alternative minimal sufficient coalitions can be
extracted directly from witness requirements:

\protect\phantomsection\label{eq-5-028}{}

\[
\mathcal{M}_{EA}(M,\Theta,\Gamma,a)
=
\min_{\subseteq}
\left\{
\operatorname{Req}_{M,\Theta,\Gamma}(w)
\;\middle|\;
\operatorname{FeasibleWitness}_{M,\Theta,\Gamma}(w,a)
\right\},
\]

where \(\min_{\subseteq}\) removes any set with a proper subset in the
family. A singleton establishes unilateral authority without excluding
other minimal coalitions. Operationally, enumerate witnesses, compute
\(\operatorname{Req}(w)\), and remove nonminimal sets; neither component
count nor a transition primitive alone determines sufficiency.

The following internal sanity results do not claim isolated mathematical
novelty.

\protect\phantomsection\label{prop-5-1}{}

\textbf{Proposition 1 (domain-collapse invariance).} Duplicating or
decomposing components within one trust domain does not create
additional independent coalition members when the selected control
assignments are unchanged. \textbf{Justification.} Projection through
\(\delta\) yields the same domain, and set formation in
\(\operatorname{Req}(w)\) removes duplicate contributions.

\protect\phantomsection\label{prop-5-2}{}

\textbf{Proposition 2 (execution-authority monotonicity).} If
\(X\subseteq Y\) and \(\mathsf{EA}_{M,\Theta,\Gamma}(X,a)\), then
\(\mathsf{EA}_{M,\Theta,\Gamma}(Y,a)\). \textbf{Justification.} The
witnessing requirement set satisfies
\(\operatorname{Req}(w)\subseteq X\subseteq Y\).

\protect\phantomsection\label{prop-5-3}{}

\textbf{Proposition 3 (minimal-coalition extraction).} Under
support-selected witness semantics, \(\mathcal{M}_{EA}\) is the
set-inclusion-minimal family of \(\operatorname{Req}(w)\) over feasible
witnesses for \(a\). \textbf{Justification.} Every execution-authority
coalition contains at least one feasible witness requirement set; a
minimal sufficient coalition therefore equals an inclusion-minimal such
set, and every inclusion-minimal requirement set is sufficient.

\subsection{5.9 Bypass, non-bypassability, and veto
coverage}\label{sec-5-9}

The run-level protected-execution predicate is:

\protect\phantomsection\label{eq-5-029}{}

\[
\mathsf{PE}(a,\rho)
\iff
a\in\mathcal{A}_{P}(\Theta)
\land
\operatorname{Occurs}(\tau_a,\rho).
\]

Run-level bypass requires a realized witness that avoids a boundary
claimed final for the same action:

\protect\phantomsection\label{eq-5-030}{}

\[
\begin{aligned}
\operatorname{BypassExec}_{M,\Theta,\Gamma}(a,B,\rho)
\iff\;&
\operatorname{ClaimedFinalFor}_{M,\Theta}(B,a)\\
&\land
\mathsf{PE}(a,\rho)\\
&\land
\exists w\;
\bigl(
\operatorname{RealizedWitness}_{M,\Theta,\Gamma}(w,a,\rho)
\land
\neg\operatorname{Traverses}(w,B)
\bigr).
\end{aligned}
\]

The predicate is witness-relative: the same run may also contain an
attempted or refused route through \(B\).

Model-level non-bypassability additionally requires a nonempty
feasible-witness set and universal traversal:

\protect\phantomsection\label{eq-5-031}{}

\[
\begin{aligned}
\operatorname{NonBypassable}_{M,\Theta,\Gamma}(B,a)
\iff\;&
\operatorname{ClaimedFinalFor}_{M,\Theta}(B,a)\\
&\land
\operatorname{HasFeasibleWitness}_{M,\Theta,\Gamma}(a)\\
&\land
\forall w\;
\bigl(
\operatorname{FeasibleWitness}_{M,\Theta,\Gamma}(w,a)
\Rightarrow
\operatorname{Traverses}(w,B)
\bigr).
\end{aligned}
\]

The existence conjunct prevents vacuity, and the universal quantifier
includes all admitted reconfiguration-enabled witnesses. Claimed-final
status records the assertion; the other conjuncts evaluate it.

Veto coverage is support-relative. For veto \(v\) governing mediated
substructure \(q\),

\protect\phantomsection\label{eq-5-032}{}

\[
\operatorname{Blocks}(v,q)
\]

holds when exercising \(v\), with other facts fixed, prevents \(q\) from
deriving \(\tau_a\). For witness \(w\),

\protect\phantomsection\label{eq-5-033}{}

\[
\operatorname{Covers}(v,w)
\]

holds when the governed mediation belongs to \(\operatorname{supp}(w)\)
and the veto invalidates it. Blocking an unused alternative or
contribution is insufficient.

The path-relative coverage set is

\protect\phantomsection\label{eq-5-034}{}

\[
\operatorname{Coverage}_{M,\Theta,\Gamma}(v,a)
=
\left\{
w
\;\middle|\;
\operatorname{FeasibleWitness}_{M,\Theta,\Gamma}(w,a)
\land
\operatorname{Covers}(v,w)
\right\}.
\]

Here \emph{path-relative} informally denotes a governed route; formal
coverage quantifies over possibly non-linear support-selected witnesses.

The global-witness-coverage predicate is:

\protect\phantomsection\label{eq-5-035}{}

\[
\begin{aligned}
\operatorname{GlobalVetoCoverage}_{M,\Theta,\Gamma}(v,a)
\iff\;&
\operatorname{HasFeasibleWitness}_{M,\Theta,\Gamma}(a)\\
&\land
\forall w\;
\bigl(
\operatorname{FeasibleWitness}_{M,\Theta,\Gamma}(w,a)
\Rightarrow
\operatorname{Covers}(v,w)
\bigr).
\end{aligned}
\]

Global coverage is non-vacuous and still requires the independence
conditions in \hyperref[sec-7]{Section 7}. These predicates are causal:
absence of recorded mediation does not prove bypass, because this paper
does not supply record-completeness or verification semantics.

\subsection{5.10 Operational procedure and illustrative end-to-end
authority derivation}\label{sec-5-10}

\phantomsection\label{alg-authority-decomposition}
\textbf{Algorithm 1: Authority-Decomposition Analysis Procedure}

\textbf{Inputs:} a concrete protected-action instance \(a\); system
configuration \(M\); threat model \(\Theta\); and environmental
assumptions \(\Gamma\).

\begin{enumerate}
\def\labelenumi{\arabic{enumi}.}
\tightlist
\item
  Identify the action-relevant protected transition \(\tau_a\) and, for
  a composite operation, separate any other consequential commit points
  into different protected actions.
\item
  Define the action-relative analytical objects: components, powers,
  controlled resources, candidate and claimed-final boundaries, vetoes,
  transition performers, and epistemic objects.
\item
  Put components in trust-domain normal form and assign each component
  through \(\delta_{M,\Theta}\); assign powers and mutable resources
  through \(\kappa_{M,\Theta}\). Record disputed assignments as
  alternative models.
\item
  Enumerate ordinary workflow powers and every update, recovery,
  override, disablement, credential-replacement, alternative-invocation,
  or other reconfiguration power admitted by \(\Theta\). Do not treat
  bypass as a primitive power.
\item
  Construct type-correct, finite, grounded, support-selected feasible
  realization witnesses for \(\tau_a\), retaining the selected AND
  inputs, OR alternative, or inclusion-minimal threshold subset.
\item
  Apply the declared reconfiguration closure: compose each admitted
  finite control certificate from \(M\) with each compatible
  successor-configuration witness, attribute the complete composite
  support to the initial model, and stop at the stated depth, state
  abstraction, or fixed point.
\item
  Compute \(\operatorname{Req}_{M,\Theta,\Gamma}(w)\) for each feasible
  witness by projecting selected discretionary support to its controller
  domains, collapsing same-domain contributions, and excluding only
  non-discretionary facts explicitly placed in \(\Gamma\).
\item
  Derive \(\mathsf{EA}_{M,\Theta,\Gamma}(X,a)\) from witness containment
  and identify unilateral and joint sufficient coalitions.
\item
  Extract \(\mathcal{M}_{EA}(M,\Theta,\Gamma,a)\) by removing every
  requirement set that has a proper subset in the feasible-witness
  requirement family.
\item
  Evaluate boundary traversal, veto coverage, I1--I4, dangerous
  authority patterns, and, separately, evidentiary independence for the
  admissible execution coalitions.
\end{enumerate}

\textbf{Outputs:} the bounded witness inventory; execution-authority and
inclusion-minimal coalitions; boundary, veto, pattern, and independence
results; and an uncertainty ledger.

The procedure may be manual or tool-assisted over a finite
representation. It makes no general discovery or scalability claim:
completeness is relative to the supplied model, power inventory,
assignments, reconfiguration closure, and witness bound. Cyclic or
extensible systems require a stated finite abstraction, depth bound, or
fixed-point rule.

\subsubsection{Illustrative End-to-End Authority
Derivation}\label{sec-5-10-illustrative-derivation}

Consider a signing appliance where \(a\) is production of signature
\(\sigma\) over digest \(h\), with key \(K\) and request identifier
\(i\). The bounded model \((a,M_0,\Theta_0,\Gamma_0)\) admits the
ordinary route plus stated firmware-replacement and
maintenance-invocation powers, but no other physical, recovery, or
hidden-interface witness. Power, availability, and usable \(K\) are
non-discretionary in \(\Gamma_0\); a discretionary key-release
controller would instead enter the requirements.

Domains \(D_C\), \(D_B\), and \(D_A\) respectively control command
production; the claimed-final boundary and ordinary signer; and update
plus maintenance signing:

\protect\phantomsection\label{eq-5-036}{}

\[
\delta(B^{*})=\delta(c_x)=D_B,
\qquad
\delta(c_u)=\delta(c_m)=\delta(c_x^{m})=D_A,
\qquad
\delta(c_g)=\delta(c_s)=D_C.
\]

The design asserts \(B^{*}\) is candidate and claimed final:

\[
\operatorname{ClaimedFinalFor}_{M_0,\Theta_0}(B^{*},a)
\land
\operatorname{CandidateFor}_{M_0,\Theta_0}(B^{*},a).
\]

\protect\phantomsection\label{tbl-5-003}{}

\emph{Table 5.3. Selected typed powers and trust-domain controllers in
the illustrative derivation.}

{\def\LTcaptype{none} 
\begin{longtable}[]{@{}
  >{\raggedright\arraybackslash}p{(\linewidth - 4\tabcolsep) * \real{0.3333}}
  >{\raggedright\arraybackslash}p{(\linewidth - 4\tabcolsep) * \real{0.3333}}
  >{\raggedright\arraybackslash}p{(\linewidth - 4\tabcolsep) * \real{0.3333}}@{}}
\toprule\noalign{}
\begin{minipage}[b]{\linewidth}\raggedright
Power
\end{minipage} & \begin{minipage}[b]{\linewidth}\raggedright
Controller
\end{minipage} & \begin{minipage}[b]{\linewidth}\raggedright
Selected effect
\end{minipage} \\
\midrule\noalign{}
\endhead
\bottomrule\noalign{}
\endlastfoot
\(p_g,p_s\) & \(D_C\) & Construct, release, and deliver the concrete
signing command. \\
\(p_b^{+}\), \(v_b\) & \(D_B\) & Accept or refuse the mediated command
at \(B^{*}\). \\
\(p_x\) & \(D_B\) & Perform \(\tau_a\) after valid boundary
acceptance. \\
\(p_u\) & \(D_A\) & Install the admitted maintenance firmware and
produce successor configuration \(M_1\). \\
\(p_m,p_x^{m}\) & \(D_A\) & Invoke maintenance signing and perform
\(\tau_a\) in \(M_1\). \\
\end{longtable}
}

The support-selected ordinary witness and the reconfiguration-enabled
witness are

\protect\phantomsection\label{eq-5-038}{}

\[
w_{\mathrm{med}}:
\{p_g,p_s,p_b^{+},p_x\}
\xRightarrow{\mathrm{AND}}
\tau_a,
\qquad
\operatorname{Traverses}(w_{\mathrm{med}},B^{*}),
\]

and

\protect\phantomsection\label{eq-5-040}{}

\[
w_{\mathrm{upd}}:
M_0\xrightarrow{p_u}M_1,
\quad
\{p_m,p_x^{m}\}
\xRightarrow{\mathrm{AND}}
\tau_a,
\qquad
\neg\operatorname{Traverses}(w_{\mathrm{upd}},B^{*}).
\]

Projection through \(\delta\) and \(\kappa\) yields

\protect\phantomsection\label{eq-5-044}{}

\[
\operatorname{Req}(w_{\mathrm{med}})=\{D_C,D_B\},
\]

and

\protect\phantomsection\label{eq-5-045}{}

\[
\operatorname{Req}(w_{\mathrm{upd}})=\{D_A\}.
\]

Hence

\protect\phantomsection\label{eq-5-046}{}

\[
\mathsf{EA}(\{D_C,D_B\},a)
\quad\text{and}\quad
\mathsf{EA}(\{D_A\},a),
\]

while neither \(D_C\) nor \(D_B\) is sufficient alone. Removing
nonminimal sets gives

\protect\phantomsection\label{eq-5-048}{}

\[
\mathcal{M}_{EA}(M_0,\Theta_0,\Gamma_0,a)
=
\bigl\{\{D_C,D_B\},\{D_A\}\bigr\}.
\]

Because \(w_{\mathrm{upd}}\) avoids \(B^{*}\),

\protect\phantomsection\label{eq-5-049}{}

\[
\neg\operatorname{NonBypassable}_{M_0,\Theta_0,\Gamma_0}(B^{*},a),
\qquad
\neg\operatorname{GlobalVetoCoverage}_{M_0,\Theta_0,\Gamma_0}(v_b,a).
\]

Thus I4 fails relative to \(U=\{D_A\}\), but the avoiding witness alone
says nothing about I1--I3. With no epistemic objects, evidentiary
independence is ``not modeled.'' The example traces the bounded
derivation from powers and domain projection to witnesses, minimal
coalitions, and boundary results.

\subsection{5.11 Analytical scope and limitations}\label{sec-5-11}

The result is conditional on the supplied model. It does not prove
implementation correctness, discover omitted interfaces or powers,
establish witness completeness, or show resistance to collusion and
common-mode failure. Uncertain shared compromise boundaries remain
uncertain even after trust-domain normalization. Results also depend on
\(\Gamma\): a discretionary choice hidden as availability understates
requirements, while treating deterministic services as discretionary
overstates them.

Composite operations still require separate protected sub-actions. The
epistemic plane identifies control over evidence production,
attestation, preservation, suppression, and official designation, but
not evidence structure, relation validity, outcome consistency, trace
completeness, verifier behavior, or conformance. Missing mediation
records therefore do not prove bypass. Subject to these limits, Sections
6 and 7 use the model to derive dangerous configurations and
independence conditions.

\section{6. Dangerous Authority Patterns}\label{sec-6}

The model reveals configurations obscured by workflow diagrams:
same-domain approvals, upstream-controlled boundary inputs,
reconfiguration witnesses, and concentrated evidence control. This
section groups them into causal concentration, boundary capture,
alternative realization, administrative collapse, and epistemic
concentration. They are derived diagnostics---not primitive authority
types, automatic vulnerability findings, or security mechanisms
\citep{anderson1972, saltzer-schroeder1975, nist-sp800-193, phillips-swiler1998}.

\subsection{6.1 From powers to configurations}\label{sec-6-1}

A power list alone is insufficient: analysis uses selected causal and
epistemic support. Fix \(M,\Theta,\Gamma,a,\mathcal{D}\) and, where
relevant, \(X,U,B,v\). Results change with the action, controller
assignments, domain partition, and admitted powers.
\(\operatorname{Controls}_{M,\Theta,\Gamma}(X,p)\) means that exercising
\(p\) has complete selected support requiring no domain outside \(X\).
Reconfiguration matters only when selected in a feasible witness or
control certificate, and bypass remains derived from a witness avoiding
the designated boundary.

\subsection{6.2 Criteria and taxonomy}\label{sec-6-2}

\emph{Dangerous} is comparative and action-relative: the topology grants
a relevant capability, defeats claimed independence, or concentrates
epistemic control. A finding must identify the coalition, selected
witness or control certificate, and consequence. Co-location, update
authority, self-produced evidence, or a threshold alone is insufficient,
and the finding denotes capability rather than exercise in every run.

\protect\phantomsection\label{tbl-6-001-main}{}

\emph{Table 6.1. Dangerous authority patterns and their principal
consequences.}

{\def\LTcaptype{none} 
\begin{longtable}[]{@{}
  >{\raggedright\arraybackslash}p{(\linewidth - 4\tabcolsep) * \real{0.3333}}
  >{\raggedright\arraybackslash}p{(\linewidth - 4\tabcolsep) * \real{0.3333}}
  >{\raggedright\arraybackslash}p{(\linewidth - 4\tabcolsep) * \real{0.3333}}@{}}
\toprule\noalign{}
\begin{minipage}[b]{\linewidth}\raggedright
Pattern
\end{minipage} & \begin{minipage}[b]{\linewidth}\raggedright
Predicate or compact trigger
\end{minipage} & \begin{minipage}[b]{\linewidth}\raggedright
Principal consequence
\end{minipage} \\
\midrule\noalign{}
\endhead
\bottomrule\noalign{}
\endlastfoot
Self-authorizing executor & \texttt{SelfAuthorizingExecutor}: the exact
authorization-controller set also satisfies \(\mathsf{EA}\). & The exact
authorization-controller set can also cause \(\tau_a\) without an
outside-domain decision. \\
Approval-execution compression & The selected approval certificate
contains the witness requirement set and no outside final restriction
remains. & The approval support consumed by the witness contains a
sufficient execution coalition. \\
Upstream-satisfied boundary & \(\operatorname{CanSatisfy}(U,B,a)\). & A
nominal boundary supplies no independently controlled favorable
condition against \(U\). \\
Mutable veto & \(\operatorname{CanWeakeningAlter}(U,B,v,a)\). & \(U\)
can reconfigure the boundary or veto in a direction that expands or
simplifies realization. \\
Veto disablement or forced non-refusal &
\(\operatorname{CanDisable}(U,B,v,a)\). & \(U\) can neutralize the
restriction while a feasible witness remains. \\
Bypassable final boundary or veto & \texttt{BypassableFinalBoundary},
\texttt{BypassableVeto}, or coalition-relative
\(\operatorname{CanBypass}\). & A feasible realization avoids a boundary
represented as final; the veto variant also requires a boundary-bound
veto claim. \\
Reconfiguration--execution coupling & A selected reconfiguration power
occurs in a minimal sufficient realization witness. & Its controller
becomes part of a minimal sufficient execution coalition. \\
Owner or administrative override collapse & An admitted administrative
reconfiguration witness yields unilateral execution authority. &
Recovery, override, key, firmware, or alternative-invocation control
makes the administrator alone sufficient. \\
Self-attesting executor & An execution coalition produces or attests its
own modeled account. & The executor controls a selected production or
attestation power over its own account. \\
Evidence-control collapse & An execution coalition can suppress all
modeled contrary accounts and exclusively designate its preferred
account. & The coalition controls the authoritative account as well as
an execution route. \\
\end{longtable}
}

The supplement retains the complete quantified catalog and interaction
matrix; the main text gives representative definitions.

\subsection{6.3 Causal authority concentration}\label{sec-6-3}

\subsubsection{6.3.1 Self-authorizing executor}\label{sec-6-3-1}

A self-authorizing executor exists when the exact domains required to
exercise a relevant authorization power are themselves sufficient for
execution. Let \(Y\subseteq\mathcal{D}\) be that exact selected
controller set:

\protect\phantomsection\label{eq-6-010}{}

\[
\begin{aligned}
\operatorname{SelfAuthorizingExecutor}_{M,\Theta,\Gamma}(Y,a)
&\\[-1mm]
\iff\;&
\exists s,p,q\;
\bigl(
\operatorname{AuthorizationCertificate}_{M,\Theta,\Gamma}(s,p,a)\\
&\hspace{12mm}\land Y=\operatorname{CReq}_{M,\Theta,\Gamma}(s)
\land \operatorname{Relevant}_{M,\Theta,\Gamma}(p,a)\\
&\hspace{12mm}\land \mathsf{authorization}\in
\operatorname{Class}_{M,\Theta,\Gamma}(p,a,q)
\land \mathsf{EA}_{M,\Theta,\Gamma}(Y,a)
\bigr).
\end{aligned}
\]

Exact-set equality prevents monotonicity from producing false positives:
an authorization controller contained in a larger coalition is not
self-sufficient when an outside final restriction remains indispensable.
Approval-execution compression is distinct and requires an approval
certificate supporting the same witness, containing its requirements,
and leaving no outside final restriction.

\subsection{6.4 Boundary dependency and capture}\label{sec-6-4}

\subsubsection{6.4.1 Upstream-satisfied boundary}\label{sec-6-4-1}

An upstream-satisfied boundary is present exactly when

\protect\phantomsection\label{eq-6-013}{}

\[
\operatorname{CanSatisfy}_{M,\Theta,\Gamma}(U,B,a)
\]

holds. The predicate supplies a complete acceptance or non-refusal
certificate \(s\) for \(B\) with

\protect\phantomsection\label{eq-6-014}{}

\[
\operatorname{CReq}_{M,\Theta,\Gamma}(s)\subseteq U.
\]

The trigger concerns complete favorable support, not evaluator location
or mutability, and is distinct from alteration, disablement, and bypass.
A mutable veto instead requires
\(\operatorname{CanWeakeningAlter}(U,B,v,a)\): a selected
reconfiguration that creates a witness, removes traversal or coverage,
enables acceptance, or reduces independent requirements. Disablement
further requires neutralizing the restriction while a feasible
realization remains.

\subsection{6.5 Alternative realization and administrative
collapse}\label{sec-6-5}

\subsubsection{6.5.1 Bypassable final boundary and
veto}\label{sec-6-5-1}

The structural boundary-level trigger is

\protect\phantomsection\label{eq-6-017}{}

\[
\begin{aligned}
\operatorname{BypassableFinalBoundary}_{M,\Theta,\Gamma}(B,a)
\iff\;&
\operatorname{ClaimedFinalFor}_{M,\Theta}(B,a)\\
&\land \operatorname{HasFeasibleWitness}_{M,\Theta,\Gamma}(a)\\
&\land \neg\operatorname{GlobalNoBypass}_{M,\Theta,\Gamma}(B,a).
\end{aligned}
\]

Where the design claims a separate veto,

\protect\phantomsection\label{eq-6-018}{}

\[
\begin{aligned}
\operatorname{BypassableVeto}_{M,\Theta,\Gamma}(v,B,a)
\iff\;&
\operatorname{BypassableFinalBoundary}_{M,\Theta,\Gamma}(B,a)\\
&\land \operatorname{VetoAt}_{M,\Theta,\Gamma}(v,B,a).
\end{aligned}
\]

The coalition-relative diagnostic remains

\protect\phantomsection\label{eq-6-019}{}

\[
\operatorname{CanBypass}_{M,\Theta,\Gamma}(U,B,a).
\]

The avoiding witness may use debug, maintenance, recovery, replacement,
or an alternative command route; no primitive bypass power is inferred,
and refusal proves only failure of the refused path.
Reconfiguration--execution coupling places a selected reconfiguration
controller in an inclusion-minimal requirement set. Administrative
override collapse is stronger: an admitted reconfiguration witness has
requirement \(\{d\}\). Titles alone establish neither result.

\subsection{6.6 Epistemic concentration}\label{sec-6-6}

A self-attesting executor is an execution coalition with complete
selected support for producing or attesting its own modeled account. It
says nothing by itself about preservation, suppression, or official
designation. Evidence-control collapse is the stronger E1 failure:

\protect\phantomsection\label{eq-6-024}{}

\[
\operatorname{ExecutionCoalition}_{M,\Theta,\Gamma}(X,a)
\land
\operatorname{CanRedefineEvidence}_{M,\Theta,\Gamma}(X,a,B,v).
\]

At system level,

\protect\phantomsection\label{eq-6-025}{}

\[
\begin{aligned}
\operatorname{EvidenceControlCollapse}_{M,\Theta,\Gamma}(a,B,v)
\iff\;&
\exists X\;
\bigl(
\operatorname{ExecutionCoalition}_{M,\Theta,\Gamma}(X,a)\\
&\hspace{15mm}\land
\operatorname{CanRedefineEvidence}_{M,\Theta,\Gamma}(X,a,B,v)
\bigr).
\end{aligned}
\]

The trigger requires contrary account possibilities, complete
suppression without indispensable outside preservation, and exclusive
official designation without an outside decision. It is not merely
``execution plus evidence authority.'' This paper identifies controllers
of five evidence functions but does not define validity, completeness,
relation correctness, consistency, verifier behavior, or conformance
\citep{w3c-prov-dm, schneier-kelsey1999, rfc5848, rfc9334}.

\subsection{6.7 Interaction and design implications}\label{sec-6-7-main}

Patterns may interact without entailing one another; the supplement
gives the full matrix. Review should derive witnesses,
\(\operatorname{Req}(w)\), and \(\mathcal{M}_{EA}\) before assigning
labels, then test each claimed-final boundary for traversal, veto
coverage, alteration, satisfaction, disablement, and bypass. Causal and
epistemic remedies remain separate. Hardware separation, thresholds,
approval, cryptography, and isolation matter through the independent
contributions they add
\citep{saltzer-schroeder1975, clark-wilson1987, shamir1979, fips140-3, pkcs11-v3.1, hse-control-systems, nist-sp800-193}.

\subsection{6.8 Scope and limitations}\label{sec-6-10}

The inventory is non-exhaustive and conditional on
\(M,\Theta,\Gamma,a\), the domain partition, and modeled support.
Presence establishes capability or dependency, not exploitation; absence
proves neither completeness nor security. Concentration may also be a
proportionate availability choice. The framework makes it visible rather
than declaring every concentrated architecture defective.

\section{7. Independence Conditions}\label{sec-7}

Physical separation does not establish authority independence when an
upstream domain controls policy, updates, recovery, credentials, or
accepted inputs; nor does current upstream insufficiency show that the
boundary causes that insufficiency. This section therefore tests whether
a designated upstream coalition can alter, satisfy, disable, or avoid a
claimed-final restriction. It separately tests whether an execution
coalition can suppress contrary accounts and exclusively designate its
preferred account.

\subsection{7.1 Independence is relative to an upstream coalition and
protected action}\label{sec-7-1}

Fix \((M,\Theta,\Gamma,a,B,U)\) and, for a veto claim, \(v\).
Independence is relative to this tuple. \emph{Upstream} is analytical:
\(U\) is the coalition whose unilateral control is tested, including
maintenance or update domains absent from the normal route. The
capability claim covers every reconfiguration, recovery, override,
disablement, and alternative invocation admitted by \(\Theta\).

A \textbf{control certificate} \(s\) applies the Section 5 support
discipline to an intermediate effect such as alteration, acceptance, or
disablement. It is finite, grounded, well-founded, and support-selected,
but does not conclude \(\tau_a\).

For a control certificate, let

\protect\phantomsection\label{eq-7-001}{}

\[
\operatorname{CReq}_{M,\Theta,\Gamma}(s)\subseteq\mathcal{D}
\]

be its required-domain set, constructed like \(\operatorname{Req}(w)\):
project selected components through \(\delta\), powers and resources
through \(\kappa\), retain selected alternatives or thresholds, collapse
same-domain contributions, exclude only non-discretionary
\(\Gamma\)-facts, and include every reconfiguration step. Thus,

\protect\phantomsection\label{eq-7-002}{}

\[
\operatorname{CReq}_{M,\Theta,\Gamma}(s)\subseteq U
\]

means that \(U\) can realize the intermediate effect without an
indispensable outside-domain decision.

Where a veto is claimed, its type and scope must be linked to the
restrictive contribution supplied through the boundary. The predicate

\protect\phantomsection\label{eq-7-003}{}

\[
\operatorname{BoundaryVetoContribution}_{M,\Theta,\Gamma}(v,B,a,q)
\]

holds when \(v\) is a veto over the \(B\)-governed substructure \(q\)
and its selected exercise supplies that boundary's restrictive outcome.
External cancellation or an unrelated kill switch is insufficient.

The typed veto claim is

\protect\phantomsection\label{eq-7-004}{}

\[
\begin{aligned}
\operatorname{VetoAt}_{M,\Theta,\Gamma}(v,B,a)
\iff\;&
\operatorname{CandidateFor}_{M,\Theta}(B,a)\\
&\land
\exists q\in\operatorname{Sub}(\mathcal{H}_{C})\;
\operatorname{BoundaryVetoContribution}_{M,\Theta,\Gamma}
(v,B,a,q).
\end{aligned}
\]

\(\operatorname{InvalidatesThrough}(v,B,w;q)\) means that exercising
\(v\), with other facts fixed, invalidates \(\operatorname{supp}(w)\)
through \(B\)'s contribution rather than another control surface. Then

\protect\phantomsection\label{eq-7-005}{}

\[
\begin{aligned}
\operatorname{CoversAt}_{M,\Theta,\Gamma}(v,B,w,a)
\iff\;&
\operatorname{FeasibleWitness}_{M,\Theta,\Gamma}(w,a)\\
&\land
\exists q\;
\bigl(
\operatorname{BoundaryVetoContribution}_{M,\Theta,\Gamma}
(v,B,a,q)\\
&\hspace{22mm}\land\;
q\subseteq\operatorname{supp}(w)\\
&\hspace{22mm}\land\;
\operatorname{InvalidatesThrough}_{M,\Theta,\Gamma}(v,B,w;q)
\bigr).
\end{aligned}
\]

The non-vacuous global predicate is

\protect\phantomsection\label{eq-7-006}{}

{\small
\[
\begin{aligned}
\begin{gathered}
\operatorname{GlobalVetoCoverageAt}_{M,\Theta,\Gamma}\\[-1mm]
(v,B,a)
\end{gathered}
\iff\;&
\operatorname{VetoAt}_{M,\Theta,\Gamma}(v,B,a)\\
&\land
\operatorname{HasFeasibleWitness}_{M,\Theta,\Gamma}(a)\\
&\land
\forall w\;
\bigl(
\operatorname{FeasibleWitness}_{M,\Theta,\Gamma}(w,a)
\Rightarrow
\operatorname{CoversAt}_{M,\Theta,\Gamma}(v,B,w,a)
\bigr).
\end{aligned}
\]
}

\subsection{7.2 Upstream insufficiency, boundary independence, and
global finality}\label{sec-7-2}

Three claims must be separated. \textbf{Upstream insufficiency} is:

\protect\phantomsection\label{eq-7-007}{}

\[
\begin{aligned}
\operatorname{UpstreamInsufficient}_{M,\Theta,\Gamma}(U,a)
\iff\;&
U\subseteq\mathcal{D}\\
&\land
\neg\mathsf{EA}_{M,\Theta,\Gamma}(U,a).
\end{aligned}
\]

It neither localizes the missing contribution nor identifies who
controls \(B\). \textbf{Boundary independence from \(U\)} means that
\(U\) cannot alter, satisfy, disable, or avoid the restrictive
contribution. \textbf{Global final-boundary independence} additionally
requires non-bypassability and, for a veto claim, coverage of every
feasible witness. These claims are not equivalent: insufficiency may
arise elsewhere, relative independence does not exclude another
coalition's bypass, and structural coverage does not prevent upstream
alteration or satisfaction.

Table 7.1 summarizes causal conditions I1--I4 and separate epistemic
condition E1.

\protect\phantomsection\label{tbl-7-001}{}

\emph{Table 7.1. Causal independence conditions I1--I4 and the separate
evidentiary condition E1.}

{\def\LTcaptype{none} 
\begin{longtable}[]{@{}
  >{\raggedright\arraybackslash}p{(\linewidth - 4\tabcolsep) * \real{0.3333}}
  >{\raggedright\arraybackslash}p{(\linewidth - 4\tabcolsep) * \real{0.3333}}
  >{\raggedright\arraybackslash}p{(\linewidth - 4\tabcolsep) * \real{0.3333}}@{}}
\toprule\noalign{}
\begin{minipage}[b]{\linewidth}\raggedright
Condition
\end{minipage} & \begin{minipage}[b]{\linewidth}\raggedright
Required result
\end{minipage} & \begin{minipage}[b]{\linewidth}\raggedright
Question answered
\end{minipage} \\
\midrule\noalign{}
\endhead
\bottomrule\noalign{}
\endlastfoot
I1 --- No unilateral alteration &
\(\neg\operatorname{CanAlter}(U,B,v,a)\) & Can \(U\) change
security-relevant decision semantics? \\
I2 --- No unilateral satisfaction &
\(\neg\operatorname{CanSatisfy}(U,B,a)\) & Can \(U\) alone supply
complete support for acceptance or non-refusal? \\
I3 --- No unilateral disablement &
\(\neg\operatorname{CanDisable}(U,B,v,a)\) & Can \(U\) neutralize the
restriction while leaving \(\tau_a\) realizable? \\
I4 --- Global no-bypass condition &
\(\operatorname{GlobalNoBypass}(B,a)\) & Does every member of a nonempty
feasible-witness set traverse \(B\)? \\
E1 --- No unilateral evidence redefinition &
\(\begin{gathered}
\operatorname{EvidentiaryIndependentFrom}\\
(a,B,v\mid X)
\end{gathered}\) & Can an
admissible execution coalition suppress contrary accounts and
exclusively designate its preferred account? \\
\end{longtable}
}

Structural traversal may remain true after neutralization if a composite
still passes through \(B\). Boundary-bound veto coverage fails when such
a composite neutralizes the veto. Neither property establishes I1 or I2;
I3 remains explicit to identify disablement and cover boundary-only
designs.

\subsection{7.3 I1 --- No unilateral alteration}\label{sec-7-3}

Alteration concerns restrictive semantics, not every administrative
change. Use \(v=\bot\) for a restrictive boundary without a separately
modeled veto. Let

\protect\phantomsection\label{eq-7-008}{}

\[
\operatorname{Reconfigures}_{M,\Theta,\Gamma}(s,M')
\]

mean that \(s\) is a support-selected control certificate for a finite
reconfiguration sequence admitted from \(M\) to successor configuration
\(M'\). Define

\protect\phantomsection\label{eq-7-009}{}

\[
\operatorname{SecurityRelevantAlter}(M,M';B,v,a)
\]

hold when a change to a decision mechanism affects at least one of:

\begin{itemize}
\tightlist
\item
  which causal realization witnesses for \(\tau_a\) are feasible;
\item
  whether a feasible witness traverses \(B\);
\item
  whether \(v\) covers a feasible witness through \(B\)'s restrictive
  contribution; or
\item
  whether a mediated attempt can obtain acceptance or non-refusal, or
  the independent-domain requirements of a relevant witness or control
  certificate change.
\end{itemize}

The unilateral-alteration diagnostic is

\protect\phantomsection\label{eq-7-010}{}

\[
\begin{aligned}
\operatorname{CanAlter}_{M,\Theta,\Gamma}(U,B,v,a)
\iff\;&
\exists s,M'\;
\bigl(
\operatorname{Reconfigures}_{M,\Theta,\Gamma}(s,M')\\
&\land
\operatorname{CReq}_{M,\Theta,\Gamma}(s)\subseteq U\\
&\land
\operatorname{SecurityRelevantAlter}(M,M';B,v,a)
\bigr).
\end{aligned}
\]

I1 is \(\neg\operatorname{CanAlter}\). The certificate includes every
contribution needed for \(M'\); irrelevant administrative changes do not
count, while rule, code, trust-key, or threshold changes may count even
if all witnesses still traverse \(B\). All reconfiguration powers
admitted by \(\Theta\), including recovery and update roots, are
considered.

Because \(\operatorname{CanAlter}\) includes strengthening changes,
dangerous weakening uses

\protect\phantomsection\label{eq-7-011}{}

\[
\begin{aligned}
\operatorname{CanWeakeningAlter}_{M,\Theta,\Gamma}(U,B,v,a)
\iff\;&
\exists s,M'\;
\bigl(
\operatorname{Reconfigures}_{M,\Theta,\Gamma}(s,M')\\
&\land
\operatorname{CReq}_{M,\Theta,\Gamma}(s)\subseteq U\\
&\land
\operatorname{SecurityRelevantAlter}(M,M';B,v,a)\\
&\land
\operatorname{WeakensFor}(M,M';B,v,a)
\bigr),
\end{aligned}
\]

where \(\operatorname{WeakensFor}\) requires at least one directional
effect: creation of a new feasible witness, loss of boundary traversal,
loss of \(\operatorname{CoversAt}\) coverage, newly obtainable
acceptance or non-refusal, or reduction of the independent-domain
requirements of a relevant support. Accordingly,

\protect\phantomsection\label{eq-7-012}{}

\[
\operatorname{CanWeakeningAlter}(U,B,v,a)
\Rightarrow
\operatorname{CanAlter}(U,B,v,a).
\]

I1 is a control-independence condition, not a test of
\(\mathsf{EA}(U,a)\) or adverse effect. Directional risk claims use
\(\operatorname{CanWeakeningAlter}\); \(\operatorname{CanAlter}\)
records control in either direction.

\subsection{7.4 I2 --- No unilateral satisfaction}\label{sec-7-4}

I2 asks who controls complete support for acceptance or non-refusal,
because an immutable boundary may accept only upstream-manufactured
facts.

Let

\protect\phantomsection\label{eq-7-013}{}

\[
\operatorname{AcceptanceCertificate}_{M,\Theta,\Gamma}(s,B,a)
\]

hold when \(s\) is a complete support-selected certificate for a
\(B\)-supplied acceptance or non-refusal relevant to at least one
feasible witness for \(a\). It retains every selected credential,
approval, signature, context, local condition, resource, threshold
participant, and decision contribution.

The satisfaction diagnostic is

\protect\phantomsection\label{eq-7-014}{}

\[
\begin{aligned}
\operatorname{CanSatisfy}_{M,\Theta,\Gamma}(U,B,a)
\iff\;&
\exists s\;
\bigl(
\operatorname{AcceptanceCertificate}_{M,\Theta,\Gamma}(s,B,a)\\
&\land
\operatorname{CReq}_{M,\Theta,\Gamma}(s)\subseteq U
\bigr).
\end{aligned}
\]

I2 is \(\neg\operatorname{CanSatisfy}\). Mechanical evaluation creates
no independent contribution when \(U\) controls every selected input; a
local key, sensor state, approval, or discretionary decision adds its
controller to \(\operatorname{CReq}(s)\). This is causal sufficiency,
not input quality, and no discretionary outside choice may be hidden in
\(\Gamma\). I1 and I2 may overlap, but an immutable rule can fail I2 and
an initially sound rule can fail I1.

\subsection{7.5 I3 --- No unilateral disablement}\label{sec-7-5}

Disablement removes, neutralizes, replaces, suspends, or forces
non-exercise of the restriction while \(\tau_a\) remains realizable.

Let

\protect\phantomsection\label{eq-7-015}{}

\[
\operatorname{DisablementCertificate}_{M,\Theta,\Gamma}
(s,M',B,v,w)
\]

hold when admitted certificate \(s\) transforms \(M\) to a
type-compatible \(M'\) and neutralizes \(B\) or \(v\) relative to
witness \(w\).

The disablement diagnostic is

\protect\phantomsection\label{eq-7-016}{}

\[
\begin{aligned}
\operatorname{CanDisable}_{M,\Theta,\Gamma}(U,B,v,a)
\iff\;&
\exists s,M',w\;
\bigl(
\operatorname{DisablementCertificate}_{M,\Theta,\Gamma}
(s,M',B,v,w)\\
&\land
\operatorname{CReq}_{M,\Theta,\Gamma}(s)\subseteq U\\
&\land
\operatorname{FeasibleWitness}_{M',\Theta,\Gamma}(w,a)\\
&\land
\operatorname{FeasibleWitness}_{M,\Theta,\Gamma}(s\circ w,a)
\bigr).
\end{aligned}
\]

For the composite witness,

\protect\phantomsection\label{eq-7-017}{}

\[
\operatorname{Req}_{M,\Theta,\Gamma}(s\circ w)
=
\operatorname{CReq}_{M,\Theta,\Gamma}(s)
\cup
\operatorname{Req}_{M',\Theta,\Gamma}(w),
\]

after domain collapse and exclusion of non-discretionary
\(\Gamma\)-facts. I3 is \(\neg\operatorname{CanDisable}\). Requiring a
feasible composite distinguishes disablement from denial of service. A
distinct avoiding witness establishes bypass but not disablement unless
it neutralizes \(B\)'s restriction. Conversely, disablement may still
traverse the component, so I3 can fail while
\(\operatorname{GlobalNoBypass}\) remains true. For \(v\neq\bot\),
neutralization entails

\protect\phantomsection\label{eq-7-018}{}

\[
\neg\operatorname{CoversAt}_{M,\Theta,\Gamma}(v,B,s\circ w,a),
\]

so the same composite refutes
\(\operatorname{GlobalVetoCoverageAt}(v,B,a)\).

\subsection{7.6 Relative and global bypass resistance}\label{sec-7-6}

The \(U\)-relative diagnostic is

\protect\phantomsection\label{eq-7-019}{}

\[
\begin{aligned}
\operatorname{CanBypass}_{M,\Theta,\Gamma}(U,B,a)
\iff\;&
\exists w\;
\bigl(
\operatorname{FeasibleWitness}_{M,\Theta,\Gamma}(w,a)\\
&\land
\operatorname{Req}_{M,\Theta,\Gamma}(w)\subseteq U\\
&\land
\neg\operatorname{Traverses}(w,B)
\bigr).
\end{aligned}
\]

This asks whether \(U\) can realize a witness avoiding \(B\). Define

\protect\phantomsection\label{eq-7-020}{}

\[
\begin{aligned}
\operatorname{RelativeBypassResistance}_{M,\Theta,\Gamma}
(B,a\mid U)
\iff\;&
\neg\operatorname{CanBypass}_{M,\Theta,\Gamma}(U,B,a).
\end{aligned}
\]

I4 is the stronger global condition:

\protect\phantomsection\label{eq-7-021}{}

\[
\begin{aligned}
\operatorname{GlobalNoBypass}_{M,\Theta,\Gamma}(B,a)
\iff\;&
\operatorname{NonBypassable}_{M,\Theta,\Gamma}(B,a).
\end{aligned}
\]

I4 requires a claimed-final \(B\), nonempty feasible-witness set, and
traversal by every admitted witness, including reconfiguration-enabled
ones. Relative resistance covers only \(U\). Global no-bypass is
structural and establishes neither immutability, independent inputs, nor
resistance to disablement.

\subsection{7.7 Composite causal-independence condition}\label{sec-7-7}

The \(U\)-relative control claim can be stated first:

\protect\phantomsection\label{eq-7-022}{}

\[
\begin{aligned}
\operatorname{BoundaryIndependentFrom}_{M,\Theta,\Gamma}
(B,v,a\mid U)
\iff\;&
U\subseteq\mathcal{D}\\
&\land
\operatorname{VetoAt}_{M,\Theta,\Gamma}(v,B,a)\\
&\land
\neg\operatorname{CanAlter}_{M,\Theta,\Gamma}(U,B,v,a)\\
&\land
\neg\operatorname{CanSatisfy}_{M,\Theta,\Gamma}(U,B,a)\\
&\land
\neg\operatorname{CanDisable}_{M,\Theta,\Gamma}(U,B,v,a)\\
&\land
\operatorname{RelativeBypassResistance}_{M,\Theta,\Gamma}
(B,a\mid U).
\end{aligned}
\]

This is a \(U\)-relative control claim; it does not exclude another
coalition's bypass or establish global veto coverage.

For a claimed-final boundary with a veto claim, define global causal
independence from \(U\) as

\protect\phantomsection\label{eq-7-023}{}

\[
\begin{aligned}
&\operatorname{CausallyIndependentFrom}_{M,\Theta,\Gamma}
(B,v,a\mid U)\\
\iff\;&
U\subseteq\mathcal{D}\\
&\land
\operatorname{ClaimedFinalFor}_{M,\Theta}(B,a)\\
&\land
\operatorname{VetoAt}_{M,\Theta,\Gamma}(v,B,a)\\
&\land
\operatorname{HasFeasibleWitness}_{M,\Theta,\Gamma}(a)\\
&\land
\operatorname{GlobalNoBypass}_{M,\Theta,\Gamma}(B,a)\\
&\land
\operatorname{GlobalVetoCoverageAt}_{M,\Theta,\Gamma}(v,B,a)\\
&\land
\neg\operatorname{CanAlter}_{M,\Theta,\Gamma}(U,B,v,a)\\
&\land
\neg\operatorname{CanSatisfy}_{M,\Theta,\Gamma}(U,B,a)\\
&\land
\neg\operatorname{CanDisable}_{M,\Theta,\Gamma}(U,B,v,a).
\end{aligned}
\]

The first two coverage predicates are structural; the negated
capabilities establish control independence from \(U\). I3 remains
explicit as a diagnostic even though boundary-bound global veto coverage
excludes the same disablement composite. Global no-bypass already
implies relative bypass resistance.

For a claimed-final execution boundary with no separately modeled veto,
define

\protect\phantomsection\label{eq-7-024}{}

\[
\begin{aligned}
&\operatorname{CausallyIndependentBoundaryFrom}_{M,\Theta,\Gamma}
(B,a\mid U)\\
\iff\;&
U\subseteq\mathcal{D}\\
&\land
\operatorname{ClaimedFinalFor}_{M,\Theta}(B,a)\\
&\land
\operatorname{HasFeasibleWitness}_{M,\Theta,\Gamma}(a)\\
&\land
\operatorname{GlobalNoBypass}_{M,\Theta,\Gamma}(B,a)\\
&\land
\neg\operatorname{CanAlter}_{M,\Theta,\Gamma}(U,B,\bot,a)\\
&\land
\neg\operatorname{CanSatisfy}_{M,\Theta,\Gamma}(U,B,a)\\
&\land
\neg\operatorname{CanDisable}_{M,\Theta,\Gamma}(U,B,\bot,a).
\end{aligned}
\]

The boundary-only form does not substitute for a claimed independent
veto, which must satisfy boundary-bound coverage. Current absence of a
command or credential is insufficient because all admitted
reconfiguration powers remain in scope.

\subsection{7.8 Relationship to execution authority}\label{sec-7-8}

Independence refines rather than replaces execution-authority analysis.
Define

\protect\phantomsection\label{eq-7-025}{}

\[
\begin{aligned}
&\operatorname{RequiresOutsideRestriction}_{M,\Theta,\Gamma}
(w,B,v,a\mid U)\\
\iff\;&
\operatorname{FeasibleWitness}_{M,\Theta,\Gamma}(w,a)\\
&\land
\operatorname{CoversAt}_{M,\Theta,\Gamma}(v,B,w,a)\\
&\land
\exists s\;
\bigl(
\operatorname{AcceptanceCertificate}_{M,\Theta,\Gamma}(s,B,a)\\
&\hspace{18mm}\land\;
\operatorname{SelectedWithin}(s,\operatorname{supp}(w))\\
&\hspace{18mm}\land\;
\operatorname{CReq}_{M,\Theta,\Gamma}(s)\nsubseteq U
\bigr).
\end{aligned}
\]

\(\operatorname{SelectedWithin}\) entails
\(\operatorname{CReq}(s)\subseteq\operatorname{Req}(w)\). The following
propositions state methodological implications and non-implications, not
a standalone theorem family.

\protect\phantomsection\label{prop-7-1}{}

\textbf{Proposition 7.1 (independent-final-veto implication).}\par
\noindent Suppose
\(\operatorname{CausallyIndependentFrom}(B,v,a\mid U)\) holds and

\protect\phantomsection\label{eq-7-026}{}

\[
\forall w\;
\bigl(
\operatorname{FeasibleWitness}_{M,\Theta,\Gamma}(w,a)
\Rightarrow
\operatorname{RequiresOutsideRestriction}_{M,\Theta,\Gamma}
(w,B,v,a\mid U)
\bigr).
\]

Then

\protect\phantomsection\label{eq-7-027}{}

\[
\neg\mathsf{EA}_{M,\Theta,\Gamma}(U,a).
\]

\textbf{Justification.} If \(\mathsf{EA}(U,a)\) held, some feasible
\(w\) would satisfy \(\operatorname{Req}(w)\subseteq U\). The selected
acceptance certificate would then satisfy
\(\operatorname{CReq}(s)\subseteq\operatorname{Req}(w)\subseteq U\),
contradicting \(\operatorname{RequiresOutsideRestriction}\).

\protect\phantomsection\label{prop-7-2}{}

\textbf{Proposition 7.2 (non-converse).} \(\neg\mathsf{EA}(U,a)\) does
not establish boundary independence from \(U\).

\textbf{Justification.} \(U\) may lack an unrelated command, target
resource, or contribution elsewhere in every current witness while
retaining unilateral authority to alter \(B\)'s rule, control every
accepted input, or disable its restrictive behavior. Upstream
insufficiency reports the absence of a complete witness requirement set
inside \(U\); it does not localize the missing contribution.

\protect\phantomsection\label{prop-7-3}{}

\textbf{Proposition 7.3 (limits of structural coverage).}
\(\operatorname{GlobalNoBypass}(B,a)\) alone establishes none of I1--I3.
\(\operatorname{GlobalVetoCoverageAt}(v,B,a)\) alone does not establish
I1, I2, or I4. Their conjunction does not establish I1 or I2, but under
reconfiguration-complete witness semantics it excludes
\(\operatorname{CanDisable}(U,B,v,a)\) for the same veto-bearing claim.

\textbf{Justification.} Every witness may traverse \(B\) while \(U\) can
change the boundary rule, manufacture complete support for acceptance,
or neutralize the veto without avoiding the boundary. Alteration and
satisfaction do not by themselves negate traversal or coverage.
Disablement differs: its feasible composite is included in the
initial-model witness inventory and, for a separately modeled veto, is
not covered through the neutralized contribution. Global veto coverage
therefore fails for that composite.

\protect\phantomsection\label{prop-7-4}{}

\textbf{Proposition 7.4 (relative bypass implication).} If
\(\operatorname{CanBypass}(U,B,a)\) holds, then \(\mathsf{EA}(U,a)\)
holds and \(\operatorname{GlobalNoBypass}(B,a)\) is false.

\textbf{Justification.} The diagnostic supplies a feasible witness with
\(\operatorname{Req}(w)\subseteq U\) and
\(\neg\operatorname{Traverses}(w,B)\), which satisfies the definition of
\(\mathsf{EA}\) and contradicts the universal traversal condition.

Physical separation supplies none of these implications when one domain
controls update, recovery, keys, accepted facts, policy, or
administration
\citep{anderson1972, saltzer-schroeder1975, fips140-3, pkcs11-v3.1, hse-control-systems, nist-sp800-193}.

\subsection{7.9 Evidentiary independence}\label{sec-7-9}

Evidentiary independence concerns the authoritative account, not whether
\(U\) defeats a causal restriction. To avoid treating every monotone
superset as relevant, define admissible execution coalitions by

\protect\phantomsection\label{eq-7-028}{}

\[
\begin{aligned}
\operatorname{ExecutionCoalition}_{M,\Theta,\Gamma}(X,a)
\iff\;&
\mathsf{EA}_{M,\Theta,\Gamma}(X,a)\\
&\land
\bigl(
X\in\mathcal{M}_{EA}(M,\Theta,\Gamma,a)\\
&\hspace{15mm}\lor\;
\operatorname{DesignatedOperationalCoalition}_{M,\Theta,\Gamma}(X,a)
\bigr).
\end{aligned}
\]

The second alternative requires explicit operational designation. The
five controller projections are:

\protect\phantomsection\label{eq-7-029}{}

\[
\eta_{\mathrm{produce}},\;
\eta_{\mathrm{attest}},\;
\eta_{\mathrm{preserve}},\;
\eta_{\mathrm{suppress}},\;
\eta_{\mathrm{official}}.
\]

Let \(\mathcal{E}^{-}_{X,a,B,v}\) denote modeled contrary or refusal
accounts and production possibilities, without asserting their validity,
completeness, or correctness. Define

\protect\phantomsection\label{eq-7-030}{}

\[
\operatorname{HasContraryAccountPotential}_{M,\Theta,\Gamma}
(X,a,B,v)
\iff
\mathcal{E}^{-}_{X,a,B,v}\neq\varnothing.
\]

\(\operatorname{CanSuppressContrary}\) requires \(X\) to prevent every
modeled contrary account from remaining available without indispensable
outside preservation. \(\operatorname{CanExclusivelyDesignate}\)
requires no indispensable outside official-designation decision.
\(\Gamma\) records environmental assumptions affecting availability or
preservation.

The evidence-redefinition diagnostic is

\protect\phantomsection\label{eq-7-031}{}

\[
\begin{aligned}
\operatorname{CanRedefineEvidence}_{M,\Theta,\Gamma}
(X,a,B,v)
\iff\;&
\operatorname{HasContraryAccountPotential}_{M,\Theta,\Gamma}
(X,a,B,v)\\
&\land
\operatorname{CanSuppressContrary}_{M,\Theta,\Gamma}(X,a,B,v)\\
&\land
\operatorname{CanExclusivelyDesignate}_{M,\Theta,\Gamma}(X,a).
\end{aligned}
\]

The relative property is

\protect\phantomsection\label{eq-7-032}{}

\[
\begin{aligned}
\operatorname{EvidentiaryIndependentFrom}_{M,\Theta,\Gamma}
(a,B,v\mid X)
\iff\;&
\operatorname{ExecutionCoalition}_{M,\Theta,\Gamma}(X,a)\\
&\land
\neg\operatorname{CanRedefineEvidence}_{M,\Theta,\Gamma}
(X,a,B,v).
\end{aligned}
\]

At system level,

\protect\phantomsection\label{eq-7-033}{}

\[
\begin{aligned}
\operatorname{EvidenceControlCollapse}_{M,\Theta,\Gamma}(a,B,v)
\iff\;&
\exists X\;
\bigl(
\operatorname{ExecutionCoalition}_{M,\Theta,\Gamma}(X,a)\\
&\hspace{16mm}\land\;
\operatorname{CanRedefineEvidence}_{M,\Theta,\Gamma}
(X,a,B,v)
\bigr),
\end{aligned}
\]

and

\protect\phantomsection\label{eq-7-034}{}

\[
\operatorname{SystemEvidentiaryIndependent}_{M,\Theta,\Gamma}(a,B,v)
\iff
\neg\operatorname{EvidenceControlCollapse}_{M,\Theta,\Gamma}(a,B,v).
\]

Production or self-attestation alone does not violate E1; collapse
requires non-vacuous contrary-account suppression plus exclusive
official designation. Evidentiary independence supports accountability
but is not a causal precondition for veto independence. This paper
identifies controllers of the five evidence powers, not record
structure, relation validity, outcome consistency, trace completeness,
verifier correctness, or conformance
\citep{w3c-prov-dm, schneier-kelsey1999, rfc5848, rfc9334}.

\subsection{7.10 Worked application}\label{sec-7-10}

The Section 5 signing model has two feasible witnesses:

\[
\begin{aligned}
\operatorname{Req}(w_{\mathrm{med}})&=\{D_C,D_B\},
&
\operatorname{Traverses}(w_{\mathrm{med}},B^{*}),\\
\operatorname{Req}(w_{\mathrm{upd}})&=\{D_A\},
&
\neg\operatorname{Traverses}(w_{\mathrm{upd}},B^{*}).
\end{aligned}
\]

\[
\neg\operatorname{GlobalNoBypass}(B^{*},a),
\qquad
\neg\operatorname{GlobalVetoCoverageAt}\\
(v_b,B^{*},a).
\]

For \(U=\{D_A\}\), both \(\operatorname{CanBypass}\) and \(\mathsf{EA}\)
hold. For \(U=\{D_C\}\), upstream insufficiency holds, yet global
independence fails because another coalition has \(w_{\mathrm{upd}}\).

\protect\phantomsection\label{tbl-7-002}{}

\emph{Table 7.2. Controlled independence variants for the
signing-appliance example.}

{\footnotesize\def\LTcaptype{none} 
\begin{longtable}[]{@{}
  >{\raggedright\arraybackslash}p{(\linewidth - 6\tabcolsep) * \real{0.2200}}
  >{\raggedright\arraybackslash}p{(\linewidth - 6\tabcolsep) * \real{0.2700}}
  >{\raggedright\arraybackslash}p{(\linewidth - 6\tabcolsep) * \real{0.2800}}
  >{\raggedright\arraybackslash}p{(\linewidth - 6\tabcolsep) * \real{0.2300}}@{}}
\toprule\noalign{}
\begin{minipage}[b]{\linewidth}\raggedright
Variant
\end{minipage} & \begin{minipage}[b]{\linewidth}\raggedright
Controlled change
\end{minipage} & \begin{minipage}[b]{\linewidth}\raggedright
Formal result
\end{minipage} & \begin{minipage}[b]{\linewidth}\raggedright
Independence conclusion
\end{minipage} \\
\midrule\noalign{}
\endhead
\bottomrule\noalign{}
\endlastfoot
\hyperref[sec-5]{Section 5} baseline & \(D_A\) installs maintenance
firmware and invokes \(c_x^{m}\); \(w_{\mathrm{upd}}\) avoids \(B^{*}\).
& \(\begin{gathered}
\operatorname{CanBypass}\\
(\{D_A\},B^{*},a);\\
\neg\operatorname{GlobalNoBypass}\\
(B^{*},a);\\
\neg\operatorname{GlobalVetoCoverageAt}\\
(v_b,B^{*},a).
\end{gathered}\) & I4 and
global structural coverage fail. \\
I1 variant & Remove the maintenance witness, but give \(D_C\) a
support-selected update certificate that rewrites \(B^{*}\)'s acceptance
rule. Every feasible witness in the variant still traverses \(B^{*}\). &
\(\begin{gathered}
\operatorname{CanAlter}\\
(\{D_C\},B^{*},v_b,a).
\end{gathered}\) & Structural traversal
does not prevent an I1 failure. \\
I2 variant & Keep \(B^{*}\) immutable, but make its acceptance depend
only on credentials, approvals, and contextual inputs controlled by
\(D_C\). & An acceptance certificate \(s\) has
\(\operatorname{CReq}(s)\subseteq\{D_C\}\), so
\(\operatorname{CanSatisfy}(\{D_C\},B^{*},a)\). & The separate boundary
component supplies no independently controlled acceptance condition. \\
I3 variant & Give \(D_C\) a control certificate \(s\) that forces
non-refusal at \(B^{*}\) while the mediated signing witness \(w\)
remains feasible; include \(s\circ w\) in the initial-model inventory. &
\(\begin{gathered}
\operatorname{CanDisable}\\
(\{D_C\},B^{*},v_b,a);\\
\neg\operatorname{GlobalVetoCoverageAt}\\
(v_b,B^{*},a).
\end{gathered}\) & I3 and
global veto coverage fail even if \(s\circ w\) still traverses
\(B^{*}\), so I4 may remain true. \\
E1 variant & Model at least one refusal-account production possibility,
then give the execution-controlling \(D_A\) domain power to suppress
every such account and exclusively designate its preferred signing
account as official. &
\(\begin{gathered}
\operatorname{CanRedefineEvidence}\\
(\{D_A\},a,B^{*},v_b).
\end{gathered}\) & E1 and
system evidentiary independence fail; this is separate from the causal
claim. \\
\end{longtable}
}

The variants separate mutable semantics (I1), upstream-controlled inputs
(I2), retained realizability after neutralization (I3), alternative
witnesses (I4), and authoritative-account control (E1).

\subsection{7.11 Scope and limitations}\label{sec-7-11}

All results are conditional on \((M,\Theta,\Gamma,a,B,v,U)\), the
trust-domain partition, and supplied witness and certificate
inventories. They prove neither non-collusion, common-mode independence,
partition correctness, implementation correctness, nor completeness;
hidden update, recovery, physical, or override powers can change the
result. Causal predicates do not establish policy merit or factual
justification, and E1 does not establish evidence validity,
completeness, verifier behavior, or conformance. Missing mediation
evidence is not bypass proof. The framework also does not require every
system to implement an independent veto; it makes the chosen allocation
and any claimed independence explicit.

\section{8. Multi-Domain Analytical Case Studies}\label{sec-8}

The framework is intended to distinguish authority configurations, not
merely to rename familiar workflow roles. This section therefore
compares analytical archetypes from AI-agent finance, enterprise payment
release, IoT control, HSM signing, and CI/CD deployment. The cases are
controlled demonstrations, not an empirical survey or evidence that
every implementation in a named domain has the modeled topology. Every
conclusion is conditional on the stated \(M,\Theta,\Gamma\),
trust-domain assignments, and feasible-witness inventory
\citep{nist-sp800-160v1r1}.

The purpose of using several domains is discriminative. Each comparison
holds visible features constant while changing an authority-relevant
fact: common versus independent control; shared versus independent
credential recovery; exclusion versus admission of a reconfiguration
power; an upstream-supplied versus locally controlled favorable
condition; or separated versus collapsed evidence control. The full case
derivations and matrices remain in the technical supplement.

\subsection{8.1 Analytical Evaluation Methodology}\label{sec-8-1}

Each case selects one concrete protected-action instance
\(a\in\mathcal{A}_{P}(\Theta)\) and its transition \(\tau_a\), fixes
\(M,\Theta,\Gamma\), normalizes components and powers to trust domains,
enumerates ordinary and admitted reconfiguration-enabled witnesses,
derives \(\operatorname{Req}(w)\) and \(\mathcal{M}_{EA}\), and then
evaluates boundary, I1--I4, pattern, and evidence-control predicates.
The evaluation uses controlled analytical contrasts: one
authority-relevant variable changes while the stated workflow shape or
causal support is held constant. The expected result is specified before
interpreting the variant, so the exercise tests whether the framework
distinguishes the intended authority change rather than merely restating
a domain label.

\protect\phantomsection\label{tbl-8-evaluation-design}{}

\emph{Table 8.1. Controlled analytical evaluation design.}

{\def\LTcaptype{none} 
\begin{longtable}[]{@{}
  >{\raggedright\arraybackslash}p{(\linewidth - 10\tabcolsep) * \real{0.1667}}
  >{\raggedright\arraybackslash}p{(\linewidth - 10\tabcolsep) * \real{0.1667}}
  >{\raggedright\arraybackslash}p{(\linewidth - 10\tabcolsep) * \real{0.1667}}
  >{\raggedright\arraybackslash}p{(\linewidth - 10\tabcolsep) * \real{0.1667}}
  >{\raggedright\arraybackslash}p{(\linewidth - 10\tabcolsep) * \real{0.1667}}
  >{\raggedright\arraybackslash}p{(\linewidth - 10\tabcolsep) * \real{0.1667}}@{}}
\toprule\noalign{}
\begin{minipage}[b]{\linewidth}\raggedright
Case
\end{minipage} & \begin{minipage}[b]{\linewidth}\raggedright
Protected action
\end{minipage} & \begin{minipage}[b]{\linewidth}\raggedright
Held constant
\end{minipage} & \begin{minipage}[b]{\linewidth}\raggedright
Changed variable
\end{minipage} & \begin{minipage}[b]{\linewidth}\raggedright
Expected discriminative result
\end{minipage} & \begin{minipage}[b]{\linewidth}\raggedright
Limitation of a simpler role/component view
\end{minipage} \\
\midrule\noalign{}
\endhead
\bottomrule\noalign{}
\endlastfoot
AI-agent finance & Post one specified transfer to the designated ledger.
& Four causal components and ordinary witness shape. & Shared control
versus independent boundary administration and lifecycle roots. &
\(\operatorname{Req}\) changes from one domain to two; the
self-authorization trigger disappears. & Counting components does not
reveal that their controllers collapse into one domain. \\
Enterprise payment release & Commit one specified payment release. &
Two-of-three approval threshold and deterministic evaluator. & Shared
credential recovery versus independent approver custody plus an outside
release domain. & One-domain support becomes pair-specific multi-domain
coalitions. & Counting approvers treats nominal principals as
independent despite common recovery. \\
IoT control & Commit one specified device command. & Nominal interlock
and ordinary mediated witness. & Excluded update, admitted diagnostic
witness, or forced non-refusal. & Boundary coverage, bypass, and
disablement separate into different results. & A component diagram shows
the same interlock in all variants and hides lifecycle-enabled
support. \\
HSM signing & Produce one specified signature with the selected key. &
HSM components, signer, and minimal execution coalition. &
Client-supplied acceptance inputs versus an indispensable HSM-domain
local condition. & I2 changes while \(\mathcal{M}_{EA}\) remains fixed.
& Hardware placement alone cannot identify who controls complete
favorable support. \\
CI/CD deployment & Commit one specified production deployment. & Causal
witnesses, requirement sets, and execution coalitions. & Independent
versus collapsed preservation, suppression, and official designation. &
E1 changes while causal execution authority does not. & A generic
``audited pipeline'' label does not allocate the distinct evidence
powers. \\
\end{longtable}
}

These are model-level demonstrations, not observations of events in
deployed systems and not empirical validation. Their evidence analysis
allocates production, attestation, preservation, suppression, and
official designation; it does not assess validity, completeness,
consistency, verification, or conformance. Uniform retrospective counts
of components, powers, candidate witnesses, reconfiguration composites,
and analyst time were not recorded during the original case construction
and are therefore unavailable; no workload or scalability metric is
inferred. The supplement exposes the complete case ledgers and final
coalition sets, but those ledgers are not presented as a substitute for
contemporaneous analyst-effort measurements.

\subsection{8.2 Positive and negative control: AI-agent
transaction}\label{sec-8-2}

Let \(a_A\) be the instruction to transfer a specified amount from
account \(\alpha\) to beneficiary \(\beta\), and let \(\tau_{a_A}\) be
posting that transfer to the designated ledger. Proposal, authorization,
command construction, and API delivery are prerequisites rather than the
transition itself. The shared-control configuration \(M_A^{S}\) and
independent-boundary configuration \(M_A^{I}\) hold functional component
count and ordinary witness shape constant. In \(M_A^{S}\), the agent,
policy service, API boundary, ledger interface, credentials, and their
relevant update or recovery roots project to \(D_A\). In \(M_A^{I}\),
the agent-side functions project to \(D_A\), while the final boundary,
transition performer, acceptance rule, and lifecycle roots project to
\(D_B\).

In the shared-control variant, all selected discretionary contributions
project to one domain:

\protect\phantomsection\label{eq-8-004}{}

\[
\operatorname{Req}_{M_A^{S}}(w_A)=\{D_A\},
\qquad
\mathcal{M}_{EA}(M_A^{S},\Theta_A,\Gamma_A,a_A)
=\bigl\{\{D_A\}\bigr\}.
\]

The authorization certificate also has
\(\operatorname{CReq}(s_A)=\{D_A\}\). The exact authorization-controller
set therefore satisfies execution authority, so the
self-authorizing-executor predicate holds. This is the negative control:
the result follows from the complete witness and shared domain
assignment, not merely from API invocation.

In the independent-boundary variant,

\protect\phantomsection\label{eq-8-006}{}

\[
\operatorname{Req}_{M_A^{I}}(w_A)=\{D_A,D_B\},
\qquad
\mathcal{M}_{EA}(M_A^{I},\Theta_A,\Gamma_A,a_A)
=\bigl\{\{D_A,D_B\}\bigr\}.
\]

Because

\protect\phantomsection\label{eq-8-007}{}

\[
\neg\mathsf{EA}_{M_A^{I}}(\{D_A\},a_A),
\]

the exact self-authorization trigger is false. \(D_A\) can propose,
authorize, assemble credentials, and initiate an attempt but cannot
cause the posting without \(D_B\); \(D_B\) performs the low-level
transition but lacks the selected request support. Under the stated
complete inventory, no upstream-controlled alteration, satisfaction,
disablement, or alternative witness is admitted, and the outside-domain
decision remains indispensable. Thus

\protect\phantomsection\label{eq-8-010}{}

\[
\operatorname{CausallyIndependentFrom}_{M_A^{I},\Theta_A,\Gamma_A}
\left(B_A,v_A,a_A\mid\{D_A\}\right).
\]

An independently controlled evidence domain preserves favorable and
refusal accounts and controls official designation, so neither
admissible execution coalition can suppress every contrary account and
exclusively designate its preferred account. The modeled E1 result is

\protect\phantomsection\label{eq-8-011}{}

\[
\operatorname{SystemEvidentiaryIndependent}_{M_A^{I},\Theta_A,\Gamma_A}
(a_A,B_A,v_A).
\]

This is the clearest positive control, not a certificate of complete
system security or a claim that the evidence domain is internally
distributed or incapable of misconduct
\citep{ap2-protocol, rfc6749, pci-dss-v4.0.1}.

\subsection{8.3 Additional controlled comparisons}\label{sec-8-3}

The enterprise case holds a two-of-three approval rule constant while
changing credential recovery and domain assignment. Shared recovery
projects the selected threshold support to one identity domain;
independent approver control plus a separate release contribution
produces pair-specific multi-domain coalitions. The difference is not
threshold arithmetic but common-control projection. The complete
derivation is retained in the supplement.

The IoT case holds the nominal interlock constant while changing the
admitted reconfiguration powers. Excluding diagnostic firmware preserves
modeled boundary coverage; admitting a firmware-enabled diagnostic
witness adds an avoiding realization without disabling the ordinary
interlock; and a separate forced-non-refusal variant disables the
restriction while retaining traversal. The comparison demonstrates that
bypass and disablement are neither primitive powers nor equivalent
outcomes.

In the HSM case, the execution witness and minimal coalition
\(\{D_C,D_H\}\) remain fixed. When the HSM accepts entirely on
client-supplied credentials and approvals,
\(\operatorname{CanSatisfy}(\{D_C\},B_H,a_H)\) holds and I2 fails.
Adding an indispensable \(D_H\)-controlled favorable condition makes
that predicate false and restores I2 without changing the coalition.
Hardware placement is therefore not the discriminating fact; selected
control of the acceptance certificate is.

The CI/CD case holds the causal witnesses and singleton execution
coalition \(\{D_C\}\) fixed while reallocating epistemic powers. In the
separated variant, an outside domain preserves contrary accounts and
another controls official designation. In the collapsed variant, \(D_C\)
controls preservation, complete suppression, and official designation,
so

\protect\phantomsection\label{eq-8-046}{}

\[
\operatorname{CanRedefineEvidence}(\{D_C\},a_D,B_D,v_D)
\]

and evidence-control collapse follow. The causal authority result is
unchanged. An ``audit log'' label is not enough; the difference lies in
who controls preservation, suppression, and designation
\citep{nist-sp800-218, slsa-v1.2}.

\subsection{8.4 Cross-case comparison}\label{sec-8-7}

\protect\phantomsection\label{tbl-8-003}{}

\emph{Table 8.2. Cross-case discriminative results under controlled
authority changes.}

{\def\LTcaptype{none} 
\begin{longtable}[]{@{}
  >{\raggedright\arraybackslash}p{(\linewidth - 8\tabcolsep) * \real{0.2000}}
  >{\raggedright\arraybackslash}p{(\linewidth - 8\tabcolsep) * \real{0.2000}}
  >{\raggedright\arraybackslash}p{(\linewidth - 8\tabcolsep) * \real{0.2000}}
  >{\raggedright\arraybackslash}p{(\linewidth - 8\tabcolsep) * \real{0.2000}}
  >{\raggedright\arraybackslash}p{(\linewidth - 8\tabcolsep) * \real{0.2000}}@{}}
\toprule\noalign{}
\begin{minipage}[b]{\linewidth}\raggedright
Controlled contrast
\end{minipage} & \begin{minipage}[b]{\linewidth}\raggedright
Feature held constant
\end{minipage} & \begin{minipage}[b]{\linewidth}\raggedright
Authority-relevant change
\end{minipage} & \begin{minipage}[b]{\linewidth}\raggedright
Formal result
\end{minipage} & \begin{minipage}[b]{\linewidth}\raggedright
Discriminative conclusion
\end{minipage} \\
\midrule\noalign{}
\endhead
\bottomrule\noalign{}
\endlastfoot
AI shared versus independent boundary & Four causal components and
ordinary witness shape & Shared versus independent administration,
update roots, and boundary decision & \(\operatorname{Req}(w_A)\)
changes from \(\{D_A\}\) to \(\{D_A,D_B\}\); self-authorization
disappears & Component count does not determine trust-domain
multiplicity or execution authority. \\
Enterprise shared versus independent approval & Same two-of-three
threshold and deterministic evaluator & Shared recovery versus
independently controlled approver powers and final release & One-domain
support versus pair-specific coalitions requiring the release domain & A
principal threshold does not by itself supply domain separation. \\
IoT update-excluded, diagnostic, and forced-non-refusal variants & Same
nominal interlock \(B_I\) & Exclusion, bypass-enabling firmware, or
forced non-refusal & Boundary coverage holds in the excluded model; an
avoiding witness defeats it in the diagnostic model; forced non-refusal
defeats the veto while traversal remains & Boundary analysis must
include admitted reconfiguration composites; bypass need not entail
disablement. \\
HSM client-satisfied versus local condition & Same HSM components,
signer, and minimal execution coalition & Addition of an indispensable
\(D_H\)-controlled favorable condition & I2 fails or holds while
\(\mathcal{M}_{EA}\) remains \(\{\{D_C,D_H\}\}\) & Coalition sufficiency
and boundary-input independence answer different questions. \\
CI/CD separated versus collapsed evidence & Same causal witnesses,
\(\operatorname{Req}(w)\), and \(\mathcal{M}_{EA}\) & Independent versus
execution-coalition control of preservation, suppression, and official
designation & Self-attestation without E1 collapse versus
evidence-control collapse & Evidentiary independence is not determined
by the causal witness. \\
\end{longtable}
}

The complete cross-case matrix, including all witness sets, boundary
predicates, I1--I4 outcomes, and evidence qualifications, appears in the
supplement.

\subsection{8.5 Findings}\label{sec-8-8}

Five bounded findings recur. Component count and principal count are
unreliable proxies for trust-domain decomposition. A low-level executor
may lack execution authority because it does not control the complete
selected support. Structural traversal does not establish resistance to
upstream alteration, satisfaction, or disablement. Reconfiguration can
add a coalition, add an alternative witness for an existing coalition,
or neutralize a restriction without adding an avoiding route. Finally,
self-attestation and evidence-control collapse remain distinct:
reallocating epistemic powers can change E1 while leaving every causal
witness unchanged.

These are analytical findings about the modeled archetypes, not
empirical validation, prevalence evidence, or proof of implementation
correctness. They show that the same predicates discriminate between
deliberately controlled variants across several application domains.

\subsection{8.6 Scope and limitations}\label{sec-8-9}

Every result remains conditional on the selected protected transition,
\(M,\Theta,\Gamma\), the trust-domain partition, and the modeled
witnesses and control certificates. An omitted debug interface, shared
administrator, recovery path, firmware key, physical-access power, or
common infrastructure dependency may add a witness or collapse a domain
distinction. Conversely, an exceptional power is not admitted merely
because it is imaginable; its basis and selected support must be stated.
Absence of a listed dangerous pattern does not prove system security.

The cases are mechanism-neutral. Hardware isolation, threshold approval,
multi-signature schemes, separate organizations, independent logs, and
software sandboxes may implement useful contributions, but their names
do not establish the predicates. What matters is whether an
outside-domain contribution is selected, whether every feasible witness
traverses the claimed boundary, and who controls alteration,
satisfaction, disablement, alternative realization, and evidentiary
designation.

\section{9. Havenlon Instantiation}\label{sec-9}

This section is a bounded, source-grounded instantiation of the
authority-decomposition framework against the documented Havenlon
architecture. It is not a product audit, vulnerability disclosure,
deployment-security assessment, certification, or validation case for
the framework. The analysis asks only what follows when the identified
sources are converted into typed components, powers, resources, trust
domains, realization witnesses, and evidence-control assignments under
the declared model. It does not establish the security of a deployed
product or close the deployment witness family. A favorable design
statement is not treated as an established property, and separately
named hardware is not treated as an independent trust domain without a
corresponding control assignment.

The principal result is deliberately qualified. In the split-control,
release-intended, open-state protocol model, the ordinary
source-supported witness has
\(\operatorname{Req}(w_{\mathrm{ord}})=\{D_L,D_U,D_S,D_A,D_X\}\). Among
the protocol witnesses enumerated from the reproducibly identified
source snapshot, certificate replacement yields the unique
inclusion-minimal known requirement set \(\{D_L,D_A,D_X\}\). Its
composed reconfiguration-and-acceptance support establishes I1 and I2
failures relative to \(D_L\) in that source-bounded model, while \(D_L\)
remains insufficient for the complete signing transition. Patent and
whitepaper materials document
\(\operatorname{ClaimedFinalFor}(B_A,a_{\mathrm{HV}})\), but deployed
global non-bypassability and boundary-bound veto coverage are
unresolved. Direct-bus, dormant-test, update, manufacturer, recovery,
and evidence-administration conclusions remain conditional or
unresolved.

\subsection{9.1 Source base and claim discipline}\label{sec-9-1}

The source base separates repository implementation behavior,
implementation-review requirements, product requirements, patents or
whitepapers, analytical inferences, explicit assumptions, and unresolved
facts. Each material statement also receives an analytical
judgment---\textbf{ESTABLISHED}, \textbf{REFUTED}, \textbf{CONDITIONAL},
or \textbf{UNRESOLVED}. These judgments state what the identified source
snapshot and model support; they are not object-level truth values and
do not establish what is deployed. The complete source hierarchy,
line-level grounding, build-state record, and source-conflict analysis
appear in the technical supplement and accompanying artifact/source
note.

The base architecture is \(M_{\mathrm{HV}}\). \(M_{\mathrm{HV}}^{R,S}\)
denotes the release-intended configuration with factory mode and debug
logging disabled and with split operational domains;
\(M_{\mathrm{HV}}^{R,C}\) is the common-control alternative. The
superscript \(S\) is a stipulated analytical assignment, not a finding
that the two MCU modules have independently controlled firmware, update,
recovery, or debug roots. \(M_{\mathrm{open}}\) denotes the runtime
state in which the mutable SaaS-lock value is not blocking. The bounded
threat models are:

\protect\phantomsection\label{tbl-9-001}{}

\emph{Table 9.1. Havenlon threat-model variants and explicitly
unresolved or excluded powers.}

{\def\LTcaptype{none} 
\begin{longtable}[]{@{}
  >{\raggedright\arraybackslash}p{(\linewidth - 4\tabcolsep) * \real{0.3333}}
  >{\raggedright\arraybackslash}p{(\linewidth - 4\tabcolsep) * \real{0.3333}}
  >{\raggedright\arraybackslash}p{(\linewidth - 4\tabcolsep) * \real{0.3333}}@{}}
\toprule\noalign{}
\begin{minipage}[b]{\linewidth}\raggedright
Threat model
\end{minipage} & \begin{minipage}[b]{\linewidth}\raggedright
Admitted powers
\end{minipage} & \begin{minipage}[b]{\linewidth}\raggedright
Deliberately unresolved or excluded from the bounded submodel
\end{minipage} \\
\midrule\noalign{}
\endhead
\bottomrule\noalign{}
\endlastfoot
\(\Theta_{\mathrm{HV}}^{\mathrm{proto}}\) & Misuse or compromise of the
Linux protocol peer, execution-user credential domain, SaaS credential
domain, and selected APP contributions; current certificate-slot and
runtime-state protocol operations; collusion among admitted domains. &
Physical injection onto the internal MCU link, undocumented credential
recovery, and firmware/debug replacement. Their exclusion is analytical,
not a sourced prevention claim. \\
\(\Theta_{\mathrm{HV}}^{\mathrm{bus}}\) & Everything in
\(\Theta^{\mathrm{proto}}\), plus physical or maintenance access to the
internal Security-facing link and the valid internal protocol material
required by the inspected source configuration. & Production enclosure
reachability and production credential custody remain unknown. \\
\(\Theta_{\mathrm{HV}}^{\mathrm{life}}\) & Everything above, plus
firmware replacement, debug or maintenance access, update-root use,
manufacturer or vendor powers, credential recovery, device recovery,
reset, key replacement, common infrastructure control, and
evidence-store administration where such powers exist. & Controller
assignments and prevention mechanisms are incomplete, so this model
supports conditional witnesses but not a closed global enumeration. \\
\end{longtable}
}

\(\Gamma_{\mathrm{HV}}\) contains only non-discretionary operating facts
needed by a feasible witness: power, selected message delivery,
incidental component availability, parser and cryptographic completion,
and secure-element session availability. Favorable Arbiter decisions,
credentials, a nonblocking lock value, owner decisions, updates,
governance approvals, evidence preservation, and official designation
remain modeled discretionary contributions
\citep{havenlon-arbiter-snapshot, havenlon-security-snapshot, havenlon-rpd0201-v0.4, havenlon-rpd0202-v1.2}.

\subsection{9.2 Protected action and system boundary}\label{sec-9-2}

The current firmware provides the strongest component-to-commit-point
support for protected signing. Let

\protect\phantomsection\label{eq-9-002}{}

\[
a_{\mathrm{HV}}=\operatorname{Sign}(i,h,K)
\in\mathcal{A}_{P}(\Theta_{\mathrm{HV}})
\]

denote the request instance \(i\) to produce a signature over the exact
digest or message \(h\) using secure-element key object \(K\). The
protected transition is

\protect\phantomsection\label{eq-9-003}{}

\[
\tau(a_{\mathrm{HV}})=\tau_{a_{\mathrm{HV}}},
\]

which occurs when the selected SE05x signing invocation returns success
with signature bytes written to the Security handler's signature buffer.
Arbiter acceptance precedes that commit point; response construction,
preservation, and official designation follow it. The action excludes
later asset transfer, SaaS receipt, evidence processing, key generation,
and slot mutation. Replacing \(K\) would define a different action
unless the source established recovery of the same key object and
authority, which it does not \citep{havenlon-security-snapshot}.

\subsection{9.3 Typed allocation}\label{sec-9-3}

The split model assigns the Linux protocol peer, user-signature
controller, SaaS-signature controller, Arbiter, and
Security/secure-element control to \(D_L,D_U,D_S,D_A,D_X\),
respectively. This normalization is action- and threat-model-relative.
Functional separation and visual spacing do not prove independent
lifecycle control.

\begin{figure}[p]
\centering
\protect\phantomsection\label{fig-9-1}
\includegraphics[height=0.70\textheight,width=0.72\textwidth,keepaspectratio]{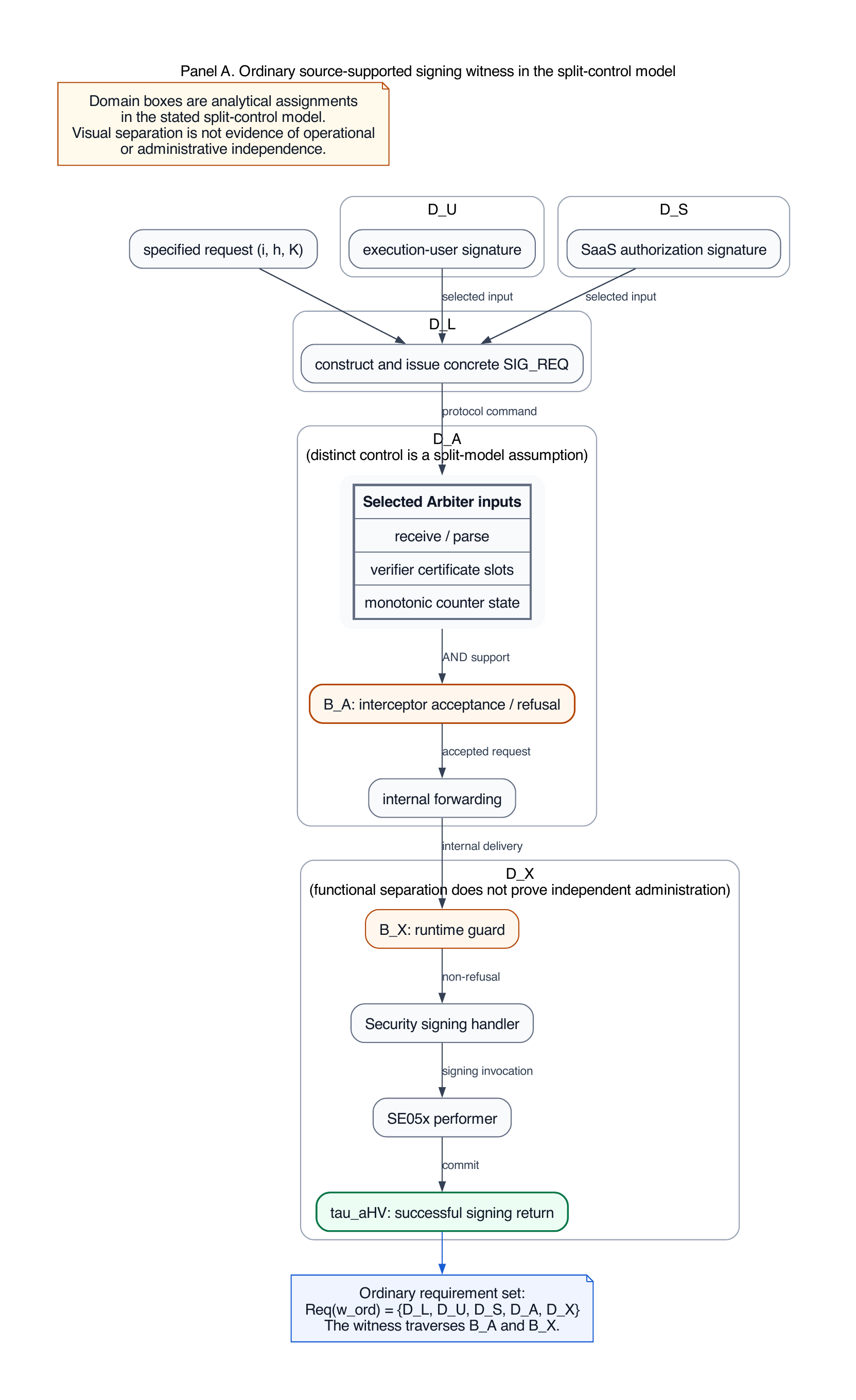}
\caption*{\textit{Figure 9.1A. Ordinary Havenlon authority model. Domain
boxes are analytical assignments under the stipulated split-control
model; spacing and hardware enclosure do not establish operational
independence.}}
\end{figure}

\begin{figure}[p]
\centering
\protect\phantomsection\label{fig-9-1b}
\includegraphics[width=\textwidth,height=0.68\textheight,keepaspectratio]{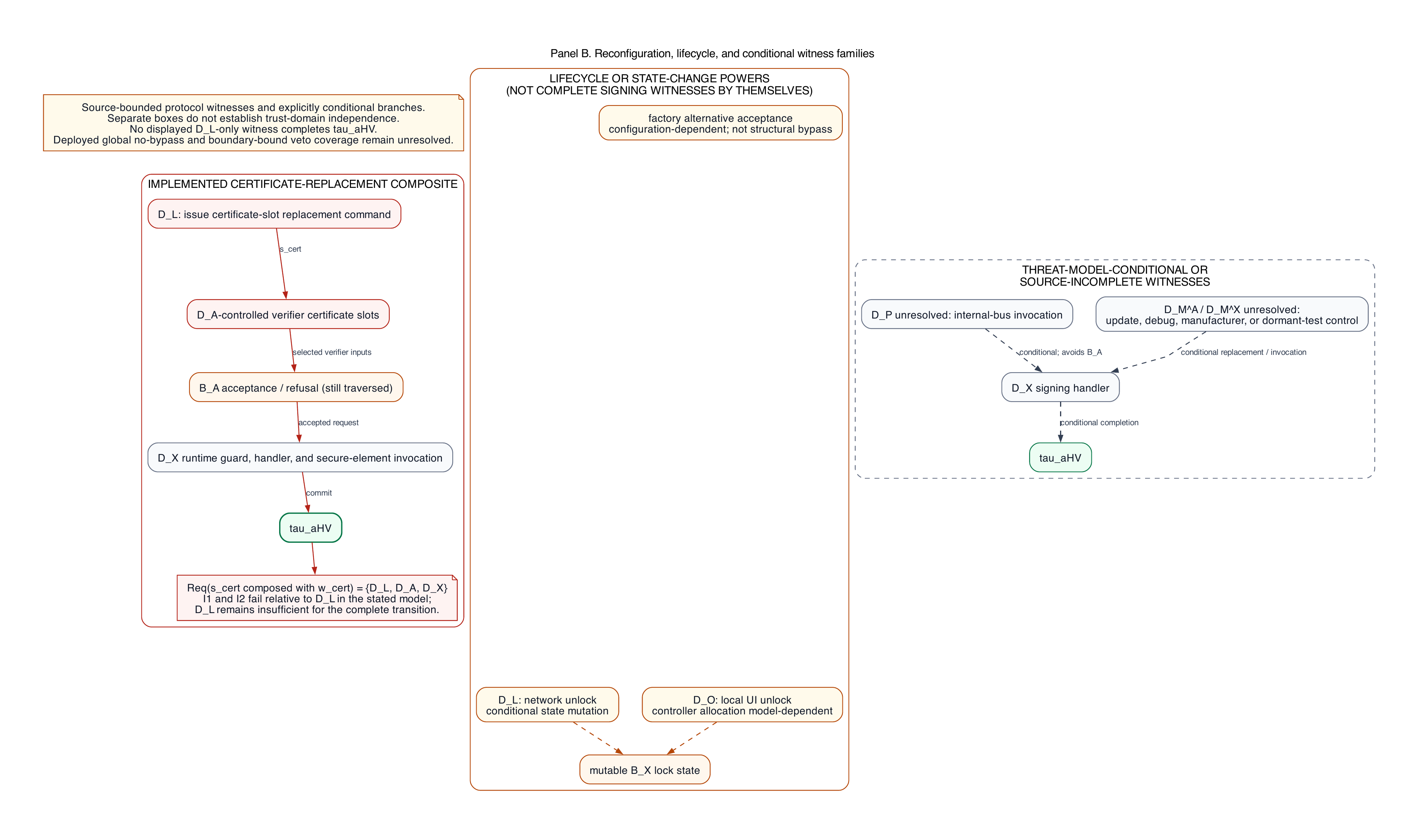}
\caption*{\textit{Figure 9.1B. Reconfiguration, lifecycle, and
conditional witnesses. The source-supported certificate-replacement
composite remains distinct from lifecycle-only state changes and
conditional or unresolved routes. No displayed Linux-domain-only route
completes the protected signing transition; deployed global
non-bypassability and boundary-bound veto coverage remain unresolved.}}
\end{figure}

The supplement provides the full component and resource allocation and
the detailed conditional bus, update, debug, manufacturer, recovery, and
dormant-test derivations behind panel B.

\subsection{9.4 Ordinary execution witness}\label{sec-9-4}

In \(M_{\mathrm{HV}}^{R,S}\cap M_{\mathrm{open}}\), the ordinary witness
selects Linux command construction, execution-user and SaaS signatures,
Arbiter parsing, counter handling, verification and forwarding, Security
non-refusal, decryption, key selection, and successful protected
signing. The complete ordinary Arbiter-acceptance certificate is
controlled by the three input domains because the Arbiter evaluates
mechanically once its selected command and signature inputs are
supplied:

\protect\phantomsection\label{eq-9-008}{}

\[
\operatorname{CReq}(s_A^{\mathrm{ord}})=\{D_L,D_U,D_S\}.
\]

The complete execution witness also requires the running Arbiter and
Security domains:

\protect\phantomsection\label{eq-9-009}{}

\[
\operatorname{Req}_{M_{\mathrm{HV}}^{R,S}}(w_{\mathrm{ord}})
=\{D_L,D_U,D_S,D_A,D_X\}.
\]

It traverses both the Arbiter and Security boundaries. Neither
acceptance establishes protected execution because forwarding, Security
processing, key selection, and successful signing remain necessary.
Conversely, refusal at the Arbiter establishes only that the refused
mediated witness does not realize the transition. Within this ordinary
split model,

\protect\phantomsection\label{eq-9-011}{}

\[
\neg\mathsf{EA}(\{D_L\},a_{\mathrm{HV}})
\quad\text{and}\quad
\neg\mathsf{EA}(\{D_X\},a_{\mathrm{HV}}).
\]

The Linux peer can initiate an attempt but cannot alone complete the
transition; the Security domain performs the low-level operation but
cannot alone supply the selected upstream support
\citep{havenlon-arbiter-snapshot, havenlon-security-snapshot}.

\subsection{9.5 Certificate replacement}\label{sec-9-5}

The most consequential implemented reconfiguration in the bounded
protocol model is certificate replacement. The Arbiter accepts a
Linux-facing certificate-write message containing a slot identifier,
user identifier, certificate bytes, and a matching digest; the inspected
handler permits the common slot range and writes the supplied
certificate. The execution-user verifier loads a caller-selected slot,
and the SaaS verifier loads fixed slot 16. No separate request-specific
approval, role check, or application-level certification-chain decision
was identified on this selected write-and-use path
\citep{havenlon-arbiter-snapshot}.

Let \(s_{\mathrm{cert}}\) replace both selected certificates with public
certificates whose private keys are controlled by the protocol peer.
Under \(\Theta_{\mathrm{HV}}^{\mathrm{proto}}\),

\protect\phantomsection\label{eq-9-013}{}

\[
\operatorname{Reconfigures}
\left(s_{\mathrm{cert}},M_{\mathrm{HV}}^{\mathrm{cert},S}\right),
\qquad
\operatorname{CReq}(s_{\mathrm{cert}})=\{D_L\}.
\]

The successor-configuration acceptance certificate likewise has
\(\operatorname{CReq}(s_{A,\mathrm{accept}}^{\mathrm{cert}})=\{D_L\}\).
The initial-model certificate must retain the reconfiguration support:

\protect\phantomsection\label{eq-9-015}{}

\[
s_A^{\mathrm{cert}}
=s_{\mathrm{cert}}\circ s_{A,\mathrm{accept}}^{\mathrm{cert}},
\]

and therefore

\protect\phantomsection\label{eq-9-017}{}

\[
\operatorname{CanSatisfy}_{M_{\mathrm{HV}}^{R,S},
\Theta_{\mathrm{HV}}^{\mathrm{proto}},\Gamma_{\mathrm{HV}}}
(\{D_L\},B_A,a_{\mathrm{HV}}).
\]

The initial-model composite realization has

\protect\phantomsection\label{eq-9-019}{}

\[
\operatorname{Req}_{M_{\mathrm{HV}}^{R,S}}
(s_{\mathrm{cert}}\circ w_{\mathrm{cert}})
=\{D_L,D_A,D_X\}.
\]

This composite still traverses the Arbiter. It is not a structural
bypass. It is a weakening alteration of the trusted-input semantics
followed by an acceptance certificate controlled by \(D_L\); it removes
\(D_U\) and \(D_S\) from the requirement set without removing \(D_A\) or
\(D_X\).

\subsection{9.6 Source-bounded coalition derivation}\label{sec-9-6}

Because the repository inspection does not establish a
deployment-complete witness inventory, define the explicitly
source-enumerated family

\protect\phantomsection\label{eq-9-033}{}

\[
\widehat{\mathcal{W}}_{\mathrm{proto}}^{\mathrm{src}}
(M,\Theta,\Gamma,a)
\]

and its inclusion-minimal requirement family

\protect\phantomsection\label{eq-9-034}{}

\[
\widehat{\mathcal{M}}_{EA,\mathrm{proto}}^{\mathrm{src}}
=\min_{\subseteq}
\left\{
\operatorname{Req}_{M,\Theta,\Gamma}(w)
\;\middle|\;
w\in\widehat{\mathcal{W}}_{\mathrm{proto}}^{\mathrm{src}}(M,\Theta,\Gamma,a)
\right\}.
\]

For the release-intended, open-state split protocol model, the
enumerated family contains \(w_{\mathrm{ord}}\) and
\(s_{\mathrm{cert}}\circ w_{\mathrm{cert}}\), with requirement sets

\protect\phantomsection\label{eq-9-036}{}

\[
\begin{aligned}
\operatorname{Req}(w_{\mathrm{ord}})
&=\{D_L,D_U,D_S,D_A,D_X\},\\
\operatorname{Req}(s_{\mathrm{cert}}\circ w_{\mathrm{cert}})
&=\{D_L,D_A,D_X\}.
\end{aligned}
\]

Hence

\protect\phantomsection\label{eq-9-037}{}

\[
\boxed{
\widehat{\mathcal{M}}_{EA,\mathrm{proto}}^{\mathrm{src}}
\left(M_{\mathrm{HV}}^{R,S},
\Theta_{\mathrm{HV}}^{\mathrm{proto}},
\Gamma_{\mathrm{HV}},a_{\mathrm{HV}}\right)
=\bigl\{\{D_L,D_A,D_X\}\bigr\}
}
\]

among the source-enumerated witnesses. This bounded minimality does not
grant unilateral authority to any of the three domains. In particular,

\protect\phantomsection\label{eq-9-038}{}

\[
\neg\exists w\in\widehat{\mathcal{W}}_{\mathrm{proto}}^{\mathrm{src}}
\;:\;
\operatorname{Req}(w)\subseteq\{D_L\}.
\]

The Linux domain remains insufficient for the complete transition. Nor
are \(D_A\) or \(D_X\) claimed globally indispensable: conditional bus,
update, debug, manufacturer, recovery, dormant-test, and other unclosed
lifecycle routes are analyzed separately in the supplement.

\subsection{9.7 Boundaries, vetoes, and independence}\label{sec-9-7}

The implemented interceptor establishes
\(\operatorname{CandidateFor}(B_A,a_{\mathrm{HV}})\). Patent and
whitepaper materials establish only the documented claim
\(\operatorname{ClaimedFinalFor}(B_A,a_{\mathrm{HV}})\), not universal
structural coverage. The ordinary Arbiter refusal is a veto at that
boundary for the mediated witness.

For \(U=\{D_L\}\) in the bounded protocol model, I1 is refuted because
\(s_{\mathrm{cert}}\) changes the trusted-key set and reduces the
witness requirement set. I2 is refuted because the composed
initial-model acceptance certificate has
\(\operatorname{CReq}(s_A^{\mathrm{cert}})=\{D_L\}\). I3 is unresolved
for the complete model: certificate replacement leaves signature
checking active, while lifecycle and debug-disablement control is
incomplete. I4 is unresolved for deployment and conditionally refuted in
\(\Theta_{\mathrm{HV}}^{\mathrm{bus}}\) when the explicitly stated
internal-link and credential assumptions hold.

I1 and I2 are already sufficient to yield

\protect\phantomsection\label{eq-9-046}{}

\[
\neg\operatorname{BoundaryIndependentFrom}
(B_A,v_A,a_{\mathrm{HV}}\mid\{D_L\})
\]

and

\protect\phantomsection\label{eq-9-047}{}

\[
\neg\operatorname{CausallyIndependentFrom}
(B_A,v_A,a_{\mathrm{HV}}\mid\{D_L\})
\]

in the bounded protocol model. These failures do not imply that \(D_L\)
can complete signing alone. Every source-enumerated protocol witness
still requires \(D_A\) and \(D_X\). Conversely, observing that the
enumerated release witnesses traverse the Arbiter is not promoted to
deployed global non-bypassability. Deployed
\(\operatorname{GlobalNoBypass}(B_A,a_{\mathrm{HV}})\) and
\(\operatorname{GlobalVetoCoverageAt}(v_A,B_A,a_{\mathrm{HV}})\) remain
\textbf{UNRESOLVED}.

\subsection{9.8 Pattern result}\label{sec-9-8}

The certificate-replacement certificate establishes an
upstream-satisfied boundary and a weakening mutable-veto result for
\(B_A\mid D_L\) in the bounded protocol model.
Reconfiguration--execution coupling is established only relative to the
source-enumerated minimal family. A direct-bus bypass is conditional on
the stronger bus threat model; the deployed bypass result remains
unresolved. Self-authorization, approval-execution compression, and
owner-only administrative collapse are not established in the
source-enumerated family and remain unresolved for the complete model.
These are predicate-backed, source-relative judgments, not product
vulnerability labels. The full ten-pattern matrix and its source bases
appear in the supplement.

\subsection{9.9 Evidence authority and E1}\label{sec-9-9}

The inspected firmware establishes Arbiter refusal/error production and
Security response/error production. Product documents assign broader
production and storage roles, but the controller map for durable
preservation, complete suppression, and exclusive official designation
is incomplete. The modeled contrary-account potential is nonempty. The
ordinary and certificate-replacement operational coalitions can produce
their own signing-response accounts and are therefore self-attesting
under the explicit account designations. That result does not entail
evidence-control collapse.

The sources establish neither complete contrary-account suppression nor
exclusive official designation for an admissible execution coalition.
Consequently,
\(\operatorname{CanRedefineEvidence}(X,a_{\mathrm{HV}},B_A,v_A)\) is
\textbf{UNRESOLVED}, and a positive E1 claim is also
\textbf{UNRESOLVED}. Causal I1/I2 failures do not decide this epistemic
result, and epistemic uncertainty does not alter the causal derivation
\citep{havenlon-rpd0301-v0.3, havenlon-rpd0302-v1.0}.

\begin{figure}[p]
\centering
\protect\phantomsection\label{fig-9-2}
\includegraphics[height=0.70\textheight,width=0.58\textwidth,keepaspectratio]{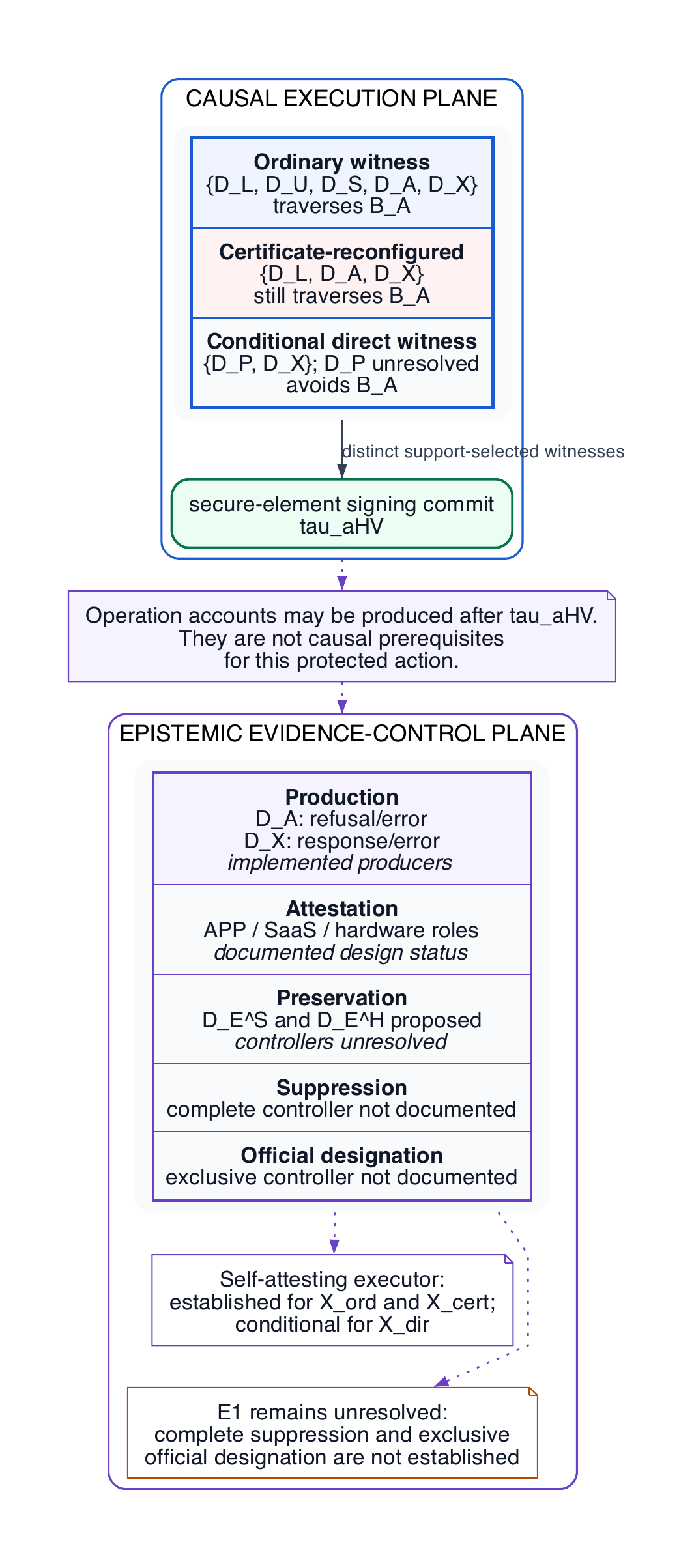}
\caption*{\textit{Figure 9.2. Causal execution support is separated from
control over evidence production, attestation, preservation,
suppression, and official designation. The figure allocates control only
and introduces no evidence-validity or verification semantics.}}
\end{figure}

\subsection{9.10 Design intent and effective authority}\label{sec-9-10}

\protect\phantomsection\label{tbl-9-005-main}{}

\emph{Table 9.2. Documented Havenlon design claims and source-bounded
authority results.}

{\def\LTcaptype{none} 
\begin{longtable}[]{@{}
  >{\raggedright\arraybackslash}p{(\linewidth - 8\tabcolsep) * \real{0.2000}}
  >{\raggedright\arraybackslash}p{(\linewidth - 8\tabcolsep) * \real{0.2000}}
  >{\raggedright\arraybackslash}p{(\linewidth - 8\tabcolsep) * \real{0.2000}}
  >{\raggedright\arraybackslash}p{(\linewidth - 8\tabcolsep) * \real{0.2000}}
  >{\raggedright\arraybackslash}p{(\linewidth - 8\tabcolsep) * \real{0.2000}}@{}}
\toprule\noalign{}
\begin{minipage}[b]{\linewidth}\raggedright
Documented design claim
\end{minipage} & \begin{minipage}[b]{\linewidth}\raggedright
Actual selected power
\end{minipage} & \begin{minipage}[b]{\linewidth}\raggedright
Controlling domain(s)
\end{minipage} & \begin{minipage}[b]{\linewidth}\raggedright
Derived authority result
\end{minipage} & \begin{minipage}[b]{\linewidth}\raggedright
Source status and remaining uncertainty
\end{minipage} \\
\midrule\noalign{}
\endhead
\bottomrule\noalign{}
\endlastfoot
Arbiter as final hardware adjudicator & Parsing, dual-signature
acceptance/refusal, and forwarding & \(D_A\) operationally; lifecycle
controller unresolved & Candidate and documented claimed-final boundary;
I1/I2 fail relative to \(D_L\) in the bounded protocol model &
Claimed-final status \textbf{ESTABLISHED as a design claim}; deployed
non-bypassability and veto coverage \textbf{UNRESOLVED} \\
Isolated Secure Executor & Low-level protected signing transition &
\(D_X\) in the split model & No \(D_X\)-only witness exists in the
source-bounded split protocol family & Source-bounded result
\textbf{ESTABLISHED}; deployed domain partition and lifecycle roots
\textbf{UNRESOLVED} \\
Joint user and SaaS authorization & Ordinary approval and authorization
signatures & \(D_U,D_S\) by analytical assignment & Both are required by
the ordinary witness; certificate replacement omits both from the
source-enumerated composite & \textbf{ESTABLISHED} within the enumerated
family; external custody and recovery remain unresolved \\
Release exclusion of factory and test modes & Build-time removal of
conditional routes & Release-build controllers unresolved & Removes
those conditional routes but not the certificate-replacement composite &
Exclusion \textbf{ESTABLISHED as a release requirement}; production
build and flags \textbf{UNRESOLVED} \\
Separate Arbiter and Security domains & Mediation and protected signing
& \(D_A,D_X\) only under the stipulated split model & Functional
separation does not establish independent lifecycle control & Functional
separation \textbf{ESTABLISHED}; independent trust domains
\textbf{UNRESOLVED} \\
Evidence Store and final receipt & Preservation and official designation
roles & Controller allocation incomplete & Typed self-attestation is
established for designated coalitions; E1 remains unresolved &
Preservation, suppression, and exclusive designation controllers
\textbf{UNRESOLVED} \\
\end{longtable}
}

The complete design/effective-authority tables, source ledger,
conditional witnesses, reproduction protocol, and blind bundle are
retained in the supplement and artifact/source layer.

\subsection{9.11 Scope and limitations}\label{sec-9-11}

This instantiation establishes only source-relative authority
conclusions for one signing action. It does not prove implementation
correctness, absence of undocumented powers, completeness of the witness
family, organizational non-collusion, common-mode independence,
correctness of the trust-domain partition, product-wide security, or the
security of any later asset transaction. The repository inspection is
not a production-image attestation. Build definitions, boot
configuration, option bytes, enclosure design, provisioned credentials,
working-tree state, and operational administration can change the
effective model.

Within those constraints, the ordinary witness in the split-control,
release-intended, open-state protocol model requires
\(\{D_L,D_U,D_S,D_A,D_X\}\). Certificate replacement yields the unique
inclusion-minimal known requirement set \(\{D_L,D_A,D_X\}\) among the
source-enumerated protocol witnesses. The composed certificate
establishes I1 and I2 failures relative to \(D_L\) in that bounded
model, although \(D_L\) remains insufficient for the complete
transition. The Arbiter's claimed-final status is documented, but
deployed global non-bypassability and boundary-bound veto coverage
remain unresolved. Direct-bus, dormant-test, update, manufacturer,
recovery, and evidence-administration results retain their conditional
or unresolved judgments. These action-, configuration-, threat-model-,
source-, and witness-relative conclusions must not be generalized to
every Havenlon operation.

\section{10. Discussion and Limitations}\label{sec-10}

The framework identifies which trust-domain coalitions can cause the
protected transition for a concrete action under a stated model. This
section interprets that conditional result without extending the formal
definitions or the bounded Havenlon findings.

\subsection{10.1 Interpreting authority decomposition as an analytical
lens}\label{sec-10-1}

Access control asks whether a request is permitted; reference-monitor
analysis asks whether a boundary mediates and enforces correctly;
credential analysis asks whether an instrument is valid; and evidence
analysis asks whether an account can be assessed under defined
semantics. Authority decomposition adds the question of which coalition
supplies complete selected causal support for the transition. It
preserves those specialized questions rather than replacing them.

The unit of analysis is the protected action, not a role or box.
Ordinary and reconfiguration-enabled witnesses are projected onto
controlling trust domains, yielding conditional causal-sufficiency
results. These results express neither permission, legitimacy, moral
responsibility, likelihood of exercise, nor attacker behavior. A
coalition may be deliberately entrusted with unilateral authority and
never exercise it; absence of a singleton does not exclude multi-domain
compromise or an omitted defect. The output is therefore action-,
model-, and threat-relative, not a permanent security label for a
product.

\subsection{10.2 What the framework reveals}\label{sec-10-2}

\subsubsection{10.2.1 Nominal roles do not determine causal
authority}\label{sec-10-2-1}

An executor may need an outside decision or resource; an administrator's
maintenance power may never complete a witness; and several approvers
may collapse to one domain through shared recovery or administration.
Conversely, upstream decision, reconfiguration, or
alternative-invocation powers may enter a sufficient coalition.
Support-selected witnesses reveal these differences without equating
component, principal, or threshold count with independent contributions.
In the HSM case, client-controlled acceptance causes I2 to fail without
granting unilateral client execution. In the Havenlon model, certificate
replacement reduces the known requirement set while leaving the Linux
domain insufficient for the complete transition.

\subsubsection{10.2.2 Component separation does not imply trust-domain
separation}\label{sec-10-2-2}

Physical, process, service, and organizational separation matters only
through control. Shared update roots, recovery, administration,
deployment, key custody, or maintenance may collapse separate boxes into
one domain; independently controlled surfaces of one device may require
analytical decomposition. Trust-domain normalization exposes this
dependence without preferring hardware or logical separation as such.

\subsubsection{10.2.3 Reconfiguration is part of execution
analysis}\label{sec-10-2-3}

Admitted updates, recovery, overrides, maintenance, disablement, and
alternative invocation belong in causal analysis. They may add a smaller
coalition, add an avoiding witness without changing minimal coalitions,
or neutralize a restriction while preserving traversal. This is why
lifecycle control, traversal, veto coverage, and I1--I4 remain distinct.
Inclusion still requires an explicit threat-model basis and controller
assignment; unresolved manufacturer, physical, debug, or recovery powers
remain conditional or unresolved, as in Section 9.

\subsubsection{10.2.4 Evidence authority is a separate analytical
dimension}\label{sec-10-2-4}

The causal topology does not determine control of the operation account.
Independent logging does not prevent execution, and an independent
execution boundary does not preserve or independently designate a
competing account. This paper identifies control of production,
attestation, preservation, suppression, and official designation, but
not truth, completeness, consistency, verification, or conformance.
Missing mediation records cannot prove bypass without a separate
completeness premise.

\subsection{10.3 Relationship to existing security
practice}\label{sec-10-3}

Authority decomposition complements existing mechanisms by making their
combined authority effect explicit. Access and usage control supply
typed decisions; reference-monitor and complete-mediation arguments
inform traversal; separation-of-duty and thresholds supply AND, OR, and
threshold structure; secure execution components supply boundary,
resource, or transition roles; and provenance or logging systems supply
evidence-control facts. The framework then asks who controls these
contributions, including lifecycle and recovery powers. It neither
replaces specialist evaluation nor collapses its conclusions into a
security score.

\subsection{10.4 Limitations}\label{sec-10-4}

\subsubsection{10.4.1 Model and witness completeness}\label{sec-10-4-1}

Hidden maintenance, updates, tests, emergency credentials, recovery,
physical access, or administrative relationships may add witnesses or
collapse domains. Positive coverage and independence claims are only as
strong as the inventories. Composite workflows require separate
protected actions, and source-bounded absence must remain unresolved
rather than becoming a negative claim.

\subsubsection{10.4.2 Threat-model and environmental
dependence}\label{sec-10-4-2}

Changes to update, recovery, manufacturer, physical, cloud, credential,
or environmental assumptions can change witnesses and coalitions.
Excluding realistic powers overstates separation; admitting unsupported
ones understates it. Comparisons require aligned actions and explicit
assumptions, even when sensitivity analysis uses justified variants.

\subsubsection{10.4.3 Trust-domain assignment}\label{sec-10-4-3}

Domain assignment is a judgment about common control and compromise, not
a naming exercise. Shared identity, administration, updates, recovery,
custody, or operations may collapse entities; joint or alternative
control must remain explicit. Disputed assignments require alternative
partitions and qualified results.

\subsubsection{10.4.4 Collusion and common-mode
failure}\label{sec-10-4-4}

Minimal coalitions establish causal sufficiency, not non-collusion or
common-mode independence. Shared personnel, infrastructure, suppliers,
incentives, governance, or vulnerability classes require separate
assurance or revised domain modeling. Coalition size is therefore not a
scalar safety score.

\subsubsection{10.4.5 Implementation and deployment
correctness}\label{sec-10-4-5}

The framework does not prove code, firmware, cryptography, hardware,
configuration, build, provisioning, or deployment correctness.
Specialist verification and testing may support model relations but
remain separate. Havenlon source findings do not attest a production
image or certify a product.

\subsubsection{10.4.6 Evidence and verification
limits}\label{sec-10-4-6}

The epistemic plane allocates five evidence powers but defines no
evidence structure, validity, provenance relations, trace completeness,
outcome consistency, verifier correctness, or conformance. Missing
traversal evidence therefore does not prove bypass. Evidentiary
independence supports accountability but does not establish causal
independence.

The principal limitations are summarized below.

\protect\phantomsection\label{tbl-10-001}{}

\emph{Table 10.1. Principal limitations and excluded inferences.}

{\def\LTcaptype{none} 
\begin{longtable}[]{@{}
  >{\raggedright\arraybackslash}p{(\linewidth - 4\tabcolsep) * \real{0.3333}}
  >{\raggedright\arraybackslash}p{(\linewidth - 4\tabcolsep) * \real{0.3333}}
  >{\raggedright\arraybackslash}p{(\linewidth - 4\tabcolsep) * \real{0.3333}}@{}}
\toprule\noalign{}
\begin{minipage}[b]{\linewidth}\raggedright
Limitation
\end{minipage} & \begin{minipage}[b]{\linewidth}\raggedright
Why it matters
\end{minipage} & \begin{minipage}[b]{\linewidth}\raggedright
What remains outside scope
\end{minipage} \\
\midrule\noalign{}
\endhead
\bottomrule\noalign{}
\endlastfoot
Model and witness completeness & Omitted powers, interfaces,
relationships, or realizations can add witnesses, change coalitions, or
defeat a coverage claim. & Automatic discovery of hidden powers; proof
that the witness inventory or graph is complete. \\
Threat-model and environmental dependence & Changes to admitted update,
recovery, physical, manufacturer, cloud, credential, or environmental
assumptions can change the authority result. & Security guarantees
against capabilities excluded from \(\Theta\); assumption-free
architecture comparisons. \\
Trust-domain assignment & Shared control or an incorrect partition can
collapse apparently independent contributions and alter minimal
coalitions. & Automatic proof of the correct organizational,
administrative, or compromise boundary. \\
Collusion and common-mode failure & Distinct modeled domains may still
collude or fail together through shared people, infrastructure,
suppliers, incentives, or vulnerabilities. & Probabilistic collusion
analysis, organizational assurance, supply-chain assurance, and
common-cause reliability. \\
Implementation and deployment correctness & A modeled relation may not
match code, firmware, hardware, build configuration, provisioning, or
the deployed artifact. & Proof of software, cryptographic, hardware,
configuration, or deployment correctness; product certification. \\
Evidence and verification limits & Control over accounts does not
establish their truth, completeness, consistency, or correct appraisal.
& Evidence validity, relation validity, trace completeness, outcome
consistency, verifier semantics, and conformance. \\
\end{longtable}
}

Accordingly, a modeled bypass is a capability claim, not evidence of
prevalence or exploitation; absence of a listed pattern does not prove
security. Concentration may be deliberate, and evidence concentration
need not satisfy evidence-control collapse.

\subsection{10.5 Design implications}\label{sec-10-5}

Remediation should follow the failed predicate: govern weakening changes
for I1, preserve an outside favorable contribution for I2, constrain
disablement and recovery for I3, constrain avoiding witnesses for I4,
and separate preservation or official designation for E1 where
appropriate. The choice remains action-, availability-, recovery-, and
governance-dependent. Design comparison should examine complete modeled
witness families and specialist mechanism evidence rather than maximize
coalition size or apply one universal prescription.

\section{Conclusion}\label{conclusion}

\subsection{1. Restating the analytical problem}\label{sec-conclusion-1}

High-risk automated systems combine authorization, delegated
credentials, specialized executors, lifecycle and recovery powers,
maintenance paths, and evidence subsystems. The nominal workflow does
not reveal which trust-domain coalitions can cause a concrete protected
transition. This paper asks that action-relative question under an
explicit model, threat model, environment, witness inventory, and domain
assignment; it does not predict attackers, detect vulnerabilities, or
certify products.

\subsection{2. Framework contribution}\label{sec-conclusion-2}

The integrated framework provides an action-relative authority
vocabulary and a typed model over components, domains, powers,
resources, boundaries, transitions, causal witnesses, and epistemic
objects. It projects support-selected witness requirements onto trust
domains to derive execution authority and inclusion-minimal sufficient
coalitions. Ordinary powers and admitted update, recovery, override,
disablement, maintenance, and alternative-invocation powers share one
analysis.

The framework separates traversal and boundary-bound veto coverage from
resistance to alteration, satisfaction, disablement, and bypass. It also
separates causal execution independence from control over evidence
production, attestation, preservation, suppression, and official
designation. These results support the patterns, demonstrations, and
Havenlon instantiation in \hyperref[sec-6]{Sections 6--9}.

The contribution is this integrated authority-centered organization and
derivation---not invention of access control, reference monitors,
complete mediation, capabilities, thresholds, hypergraphs, trusted
execution, HSMs, provenance, or secure logging.

\subsection{3. Main analytical insights}\label{sec-conclusion-3}

Five bounded insights follow. Execution authority follows complete
selected support, not roles. Component, principal, and threshold counts
do not establish trust-domain multiplicity. Admitted lifecycle powers
may change coalitions, add witnesses, alter restrictions, or enable
avoidance. Structural traversal alone does not establish resistance to
upstream alteration, satisfaction, or disablement. Evidence authority
remains separate: self-attestation need not be evidence-control
collapse, and independent evidence does not constrain execution.

\subsection{4. Analytical demonstrations and Havenlon
qualification}\label{sec-conclusion-4}

The five Section 8 cases demonstrate discrimination among controlled
authority variants; they are not an empirical benchmark, prevalence
study, or validation.

In the split-control, release-intended, open-state, source-bounded
Havenlon protocol model, the ordinary witness requires

\protect\phantomsection\label{eq-conclusion-001}{}

\[
\{D_L,D_U,D_S,D_A,D_X\}.
\]

Among source-enumerated protocol witnesses, certificate replacement
yields the unique inclusion-minimal known requirement set

\protect\phantomsection\label{eq-conclusion-002}{}

\[
\{D_L,D_A,D_X\}.
\]

Certificate replacement establishes I1 and I2 failures relative to
\(D_L\) within that model, while \(D_L\) remains insufficient for the
complete transition. The Arbiter's claimed-final status is a documented
design claim; deployed global non-bypassability and boundary-bound veto
coverage remain unresolved. These are qualified authority conclusions,
not security, vulnerability, or certification claims.

\subsection{5. Limits of the conclusions}\label{sec-conclusion-5}

Results remain conditional on the action, \(M,\Theta,\Gamma\),
inventories, domain assignments, and source basis. Hidden powers,
relationships, or partition errors can change them. The analysis proves
neither implementation or product security, completeness, organizational
or common-mode independence, nor any scalar safety score. It also does
not establish evidence validity, relation validity, trace completeness,
outcome consistency, verifier correctness, or conformance.

\subsection{6. Closing synthesis}\label{sec-conclusion-6}

The central result is therefore derivable rather than label-based: which
trust-domain coalitions can make the protected action happen?

\section*{Statements and Declarations}\label{statements-declarations}
\addcontentsline{toc}{section}{Statements and Declarations}

\subsection*{Conflicts of interest}\label{declaration-conflicts}
\addcontentsline{toc}{subsection}{Conflicts of interest}

All authors are affiliated with Chengdu Havenlon Security Technology
Co., Ltd., whose system is analyzed in Section 9. The authors declare no
other potential conflicts of interest with respect to the research,
authorship, or publication of this article.

\subsection*{Funding}\label{declaration-funding}
\addcontentsline{toc}{subsection}{Funding}

This work was supported by Chengdu Havenlon Security Technology Co.,
Ltd.

\subsection*{Ethics approval and consent to
participate}\label{declaration-ethics}
\addcontentsline{toc}{subsection}{Ethics approval and consent to
participate}

Not applicable. This study did not involve human participants, human
data, animals, or personally identifiable information.

\subsection*{Consent for publication}\label{declaration-consent}
\addcontentsline{toc}{subsection}{Consent for publication}

Not applicable.

\subsection*{Author contributions}\label{declaration-contributions}
\addcontentsline{toc}{subsection}{Author contributions}

Mengting Wu: Conceptualization, Methodology, Formal analysis, Writing -
original draft, Project administration. Lin Wang: Validation, Technical
review, Writing - review and editing. Yong Zhang: Supervision, Technical
review, Writing - review and editing.

\subsection*{Data and artifact
availability}\label{declaration-artifacts}
\addcontentsline{toc}{subsection}{Data and artifact availability}

The ordinary literature sources are cited in the reference list. The
source material supporting the Havenlon instantiation includes
anonymized implementation snapshots and design documents that are not
distributed with this preprint. A complete lawful public or controlled
scholarly-access route has not yet been approved. The current
blind-reproduction package is partial, and no independent reproduction
is claimed. The Havenlon section must therefore be read only as the
bounded, source-grounded instantiation described at the beginning of
Section 9.

\bibliography{paper-a-arxiv-references-v0.3.bib}

\end{document}